\documentclass[preprintnumbers, prd, showpacs, floatfix,onecolumn, superscriptaddress,nofootinbib]{revtex4}
\usepackage{amsmath}
\usepackage{latexsym}
\usepackage{graphicx,epsfig,color}
\usepackage{amssymb}
\usepackage{subfigure}
\usepackage{epstopdf}
\usepackage[colorlinks=true,citecolor=blue,linkcolor=blue,urlcolor=black]{hyperref}
\begin{document}

\title{Critical behavior of Gauss-Bonnet black holes via an alternative phase space}
\author{H. Yazdikarimi$^{1}$, A. Sheykhi$^{1,2}$\footnote{
asheykhi@shirazu.ac.ir} and Z. Dayyani$^{1}$}

\affiliation{Physics Department and Biruni Observatory, College of
Sciences, Shiraz University, Shiraz 71454, Iran\\
$^2$ Research Institute for Astronomy and Astrophysics of Maragha
(RIAAM), P. O. Box: 55134-441, Maragha, Iran}

\begin{abstract}
Recently, it was argued that charged Anti-de Sitter (AdS) black
holes admit critical behavior, without extending phase space,
similar to the Van der Waals fluid system in the $Q^2-\Psi$ plans
where $\Psi=1/v$ (the conjugate of $Q^2$) is the inverse of the
specific volume \cite{Dehy}. In this picture, the square of the
charge of the black hole, $Q^2$, is treated as a thermodynamic
variable and the cosmological constant $\Lambda$ is fixed. In this
paper, we would like to examine whether this new approach toward
critical behaviour of AdS black holes can work in other gravity
such as Gauss-Bonnet (GB) gravity as well as in higher dimensional
spacetime. We obtain the equation of state, $Q^2=Q^2(\Psi, T)$,
Gibbs free energy and the critical quantities of the system, and
study the effects of the GB coupling $\tilde{\alpha}$ on their
behaviour. We find out that the critical quantities have
reasonable values, provided the GB coupling constant,
$\tilde{\alpha}$, is taken small and the horizon topology is
assumed to be $(d-2)$-sphere. Finally, we calculate the critical
exponents and show that they are independent of the model
parameters and have the same values as the Van der Waals system
which is predicted by the mean field theory.
\end{abstract}

\maketitle

\section{Introduction}
Thermodynamics of black holes has been started around five decades
ago in $1970's$ by the works of Hawking and Bekenstein. Since the
discovery of black holes thermodynamics, a lot of investigations
have been carried out to disclose the similarity between the laws
of black holes mechanics and the usual thermodynamical systems on
the earth. The motivation for this investigation is to understand
the microscopic structure of black holes and hence shed light on
the quantum theory of gravity as well. Thermodynamics of charged
black holes in the background of asymptotically AdS spacetimes is
of specific interest, mainly due to the duality between gravity in
AdS spacetime and the Conformal Field Theory (CFT) living on its
boundary. According to AdS/CFT correspondence
\cite{Maldacena,Gubser,Witten}, thermodynamics of black holes in
an AdS space can be recognized by that of dual strong coupled CFT
on the boundary of the AdS spacetime. Besides, it has been shown
that there is a complete analogy between charged black holes in
AdS space and the Van der Waals liquid-gas system with their
critical exponents coincide with those of the Van der Waals system
which is predicted by the mean field theory. In this picture, the
phase space of black holes thermodynamics is extended such that
the cosmological constant is regarded as the thermodynamic
pressure and its conjugate quantity as a thermodynamic volume
\cite{Do1,Kastor,Do2,Do3,Ce1,Ur,Mann1}. Interestingly enough, it
has been displayed that both systems have extremely similar phase
diagrams \cite{Mann1}. This analogy has been generalized to higher
dimensional charged black holes \cite{Mann2}, rotating black holes
\cite{Altam,Sherkat,Sherkat1} and dilaton black holes
\cite{Kamrani}. The studies were also enlarged to the critical
behavior of nonlinear black holes \cite{Nonlinear}. When the gauge
field is in the form of Born-Infeld nonlinear electrodynamics, one
should extend the phase space and introduce a new thermodynamic
quantity conjugate to the Born-Infeld parameter which is necessary
for consistency of the first law of thermodynamics as well as the
corresponding Smarr relation \cite{MannBI, Dayyani1}.

It is also of great interest to consider the higher curvature
corrections to the Einstein gravity. In these theories the entropy
expressions are not proportional to the area of the horizon, and
instead are given by a more complicated relation depending on
higher-curvature terms \cite{Iyer}. One of the most important and
useful batch of these kind of theories are the Lovelock gravity
theories \cite{Lov}, which lead to second order differential
equations for the metric functions. The second-order Lovelock
theory of gravity is well-known as GB gravity, which contains
higher curvature terms in the action. The phase structure of GB
black holes in AdS spaces has been explored in \cite{Cai1, Dey}.
Motivated by the idea that the cosmological constant can be
regarded as a thermodynamic variable, critical behavior of charged
topological GB black holes in $d$-dimensional AdS spacetime has
been studied in \cite{Shao,Cai2,Zou}. Thermodynamic analogy
between a charged GB-AdS black hole and a Van der Waals liquid gas
system has been confirmed and it was shown that the result
drastically depend on $\alpha$ and dimensions of the spacetime. It
was shown that when one treats the GB coupling constant as a free
thermodynamic variable \cite{Xu}, the Van der Waals behavior is
occurred, and criticality and reentrant behaviour are observed
\cite{Wei}. Furthermore, the phase structure of asymptotically AdS
black holes in Lovelock gravity have also been explored
\cite{Lovelock}.

In all works mentioned above, the cosmological constant is
regarded as thermodynamic pressure which can vary. Although, there
are some motivations to consider the cosmological constant as a
variable, but it is more reasonable to keep it as a constant
parameter. For example, in general relativity the cosmological
constant is usually considered as a constant related to the zero
point energy of the vacuum. Motivated by the argument given in
\cite{Dehy}, we want to study the critical behavior of GB black
hole via an alternative viewpoint, in which we keep the
cosmological constant as a constant parameter and instead treat
the charge of the black hole (or more precisely $Q^2$) as an
external variable which can vary. The advantages of this approach
is that it provides  more attractive and straightforward results.
Phase structure and critical behavior of BI black holes in an AdS
space, where the charge of the system can vary and the
cosmological constant (pressure) is fixed have been investigated
in \cite{Dehy2}. It was shown the system indeed admits a reentrant
phase transition. Recently, it was shown that this method also
work for investigating the critical behaviour of Lifshitz dilaton
black holes \cite{Dayani2}, which further supports the viability
of this new approach.

This paper is organized as follows. In the next section we study
the critical behavior of $d$-dimensional charged AdS black hole
using an alternative phase space. In section \ref{Structure} we
review the solution of GB black holes and their thermodynamic
features. In section \ref{d5}, we investigate the $(Q^2-\Psi)$
phase space of the GB black holes in five dimensions and obtain
the critical quantities. Also, we will calculate the critical
exponents and Gibbs free energy of the system. In section
\ref{arbitrary d} we generalize our study to higher dimension by
investigating the critical behavior of GB black holes via an
alternative method. We use the results of section \ref{Field} to
study the accuracy of our calculations in the limit of $\alpha=0$.
The last section is devoted to the summery and conclusions.
\section{Critical behaviour of AdS
black holes in higher dimensions}\label{Field}

As we mentioned before, an alternative approach towards
investigating the critical behaviour of AdS black holes was
suggested without extending the phase space  \cite{Dehy}. The
authors of \cite{Dehy} completed the analogy between charged AdS
black holes in four dimensions and Van der Waals fluid system by
treating the square of the charge of the black hole, $Q^2$, as the
thermodynamic variable and keeping the cosmological constant
fixed. Our aim here is to generalize this new approach to higher
dimensional charged AdS black holes. The motivations is to check
whether this approach does work in higher dimensions or it is only
valid in four-dimensions. Besides, this investigation is of great
importance since it provides the background for our calculations
in the next section, where we would like to investigate the
critical behaviour of GB black holes in all higher dimensions.
\subsection{Critical behaviour of black holes in $d$-dimensions}
The action of Einstein-Maxwell theory in the background of AdS
spacetime in $d$-dimensions is given by
\begin{eqnarray}\label{action}
S &=&-\frac{1}{16\pi }\int d^{d}x\sqrt{-g}\left( R
-2\Lambda-F_{\mu \nu} F^{\mu \nu}\right),
\end{eqnarray}
where $R$ is the Ricci scalar, $\Lambda=-(d-1)(d-2)/2l^2$ is the
cosmological constant, $F_{\mu
\nu}=\partial_{\mu}A_{\nu}-\partial_{\nu}A_{\mu}$ is the
electrodynamic field tensor with $A_{\mu}$ is the gauge potential
\cite{Bril1,Cai3}
\begin{eqnarray}\label{function}
A_{t}=\frac{Q}{(d-3)r^{d-3}}.
\end{eqnarray}
The most general $d$-dimensional static metric with constant
curvature boundary may be written as
\begin{equation}
ds^{2}=-f(r)dt^{2}+\frac{dr^{2}}{f\left( r\right) }+r^{2} d
\Sigma^2_{d-2},  \label{metric}
\end{equation}%
where $d\Sigma^2_{d-2}$ stands for the line elements of a
$(d-2)$-dimensional hypersurface with constant scalar curvature
$\left( d-2\right) \left( d-3\right) k$ \ and volume $\omega
_{d-2}$. Here $k$ is a constant and characterizes the curvature of
the hypersurface. Without loss of generality, one can take $k=0,
1, -1$, such that the black hole horizon or cosmological horizon
in (\ref{metric}) can be a zero (flat), positive (elliptic) or
negative (hyperbolic) constant curvature hypersurface. The
function $f(r)$ is given by \cite{Bril1,Cai3}
\begin{eqnarray}\label{f}
f(r)=k-\frac{m}{r^{d-3}}+\frac{2Q^2}{(d-2)(d-3)r^{2(d-3)}}+\frac{r^2}{l^2},
\end{eqnarray}
where $m$ and $Q$ are, respectively, the mass and the charge
parameters which are related to the total mass and charge of the
black hole via
\begin{eqnarray}\label{mq}
M=\frac{m(d-2)}{16\pi}\omega_{d-2},\  \  \quad
\mathcal{Q}=\frac{Q}{4\pi} \omega_{d-2}.
\end{eqnarray}
The horizon radius $r_{+}$ of the black hole is the largest real
root of Eq. $f(r_{+})=0$. Taking into account the energy formation
of the thermodynamic system, it was argued that the mass of AdS
black hole, $M$, is indeed the enthalpy $H$ \cite{Kastor}. It is a
matter of calculations to show that in terms of the horizon radius
the mass is given by
 \begin{eqnarray}\label{M}
M=\frac{(d-2)\omega_{d-2} r_{+}^{d-1}}{16\pi
l^2}+\frac{k(d-2)\omega_{d-2}
r_{+}^{(d-3)}}{16\pi}+\frac{\omega_{d-2} Q^2}{8\pi (d-3)
r_{+}^{d-3}}.
\end{eqnarray}
The Hawking temperature of the black hole on the event horizon
$r_{+}$ can be calculated as
 \begin{eqnarray}\label{T}
T=\frac{f'(r_{+})}{4\pi}=\frac{(d-1)r_{+}}{4\pi
l^2}+\frac{(d-3)k}{4\pi r_{+}}-\frac{Q^2}{2\pi (d-2)
r_{+}^{2d-5}}.
\end{eqnarray}
The entropy and electric potential $\Phi$ of the black hole are
given by \cite{MannBI}
\begin{eqnarray}
&&S=\frac{r_{+}^{d-2}}{4}\omega_{d-2}\\
&&\Phi=\frac{Q}{(d-3)r_+ ^{d-3}}.
\end{eqnarray}
According to \cite{MannBI}, in the extended phase space towards
study the critical behaviour of AdS black holes, the cosmological
constant which is interpreted as a thermodynamic pressure $P$, and
its conjugate quantity, the thermodynamic volume, are given by
\begin{eqnarray}
P=-\frac{\Lambda}{8\pi}=\frac{(d-1)(d-2)}{16\pi l^2}, \  \  \ V=
\left (\frac{\partial M}{\partial
P}\right)_{Q,S}=\frac{\omega_{d-2} r_{+}^{d-1}}{d-1}.\label{pV}
\end{eqnarray}
It was shown that all these quantities satisfy the following Smarr
formula \cite{MannBI}
\begin{equation}
M=\frac{d-2}{d-3}T S+\Phi \mathcal{Q}-\frac{2}{d-3}V P,
\end{equation}
Besides, the first law of thermodynamics with variable $P$ and
fixed $\mathcal{Q}$ is written as
\begin{equation}
dM=TdS+\Phi d\mathcal{Q}+ V dP.
\end{equation}
It was shown that charged AdS black holes represent a critical
behavior similar to Van der Waals fluid, if one treat the
cosmological constant as a thermodynamic variable
\cite{Mann1,MannBI}. Although this idea has got a lot of interests
in the literatures, it was argued that by keeping the cosmological
constant as a fixed parameter and instead considering $Q^2$ as a
thermodynamic variable,  the critical behavior can be seen in
$Q^2-\Psi$ plane \cite{Dehy}. Following \cite{Dehy}, we replace
the term $\Phi dQ$ in the first law with $\Psi dQ^2$,
\begin{eqnarray}
dM=TdS+\Psi dQ^2+VdP
\end{eqnarray}
It is a matter of calculation to show that the Smarr formula takes
the form
\begin{equation}
M=\frac{d-2}{d-3}T S+\Psi Q^2-\frac{2}{d-3}V P,
\end{equation}
where we have defined
 \begin{eqnarray}
\Psi=\left (\frac{\partial M}{\partial
Q^2}\right)_{P,S}=\frac{\omega_{d-2}}{8\pi(d-3)r_{+}^{d-3}}.\label{Psi}
\end{eqnarray}
\subsection{Critical behavior}
\begin{figure}
 \centering \subfigure[$k=1$, $d=5$, $l=1$]
{\includegraphics[scale=0.3]{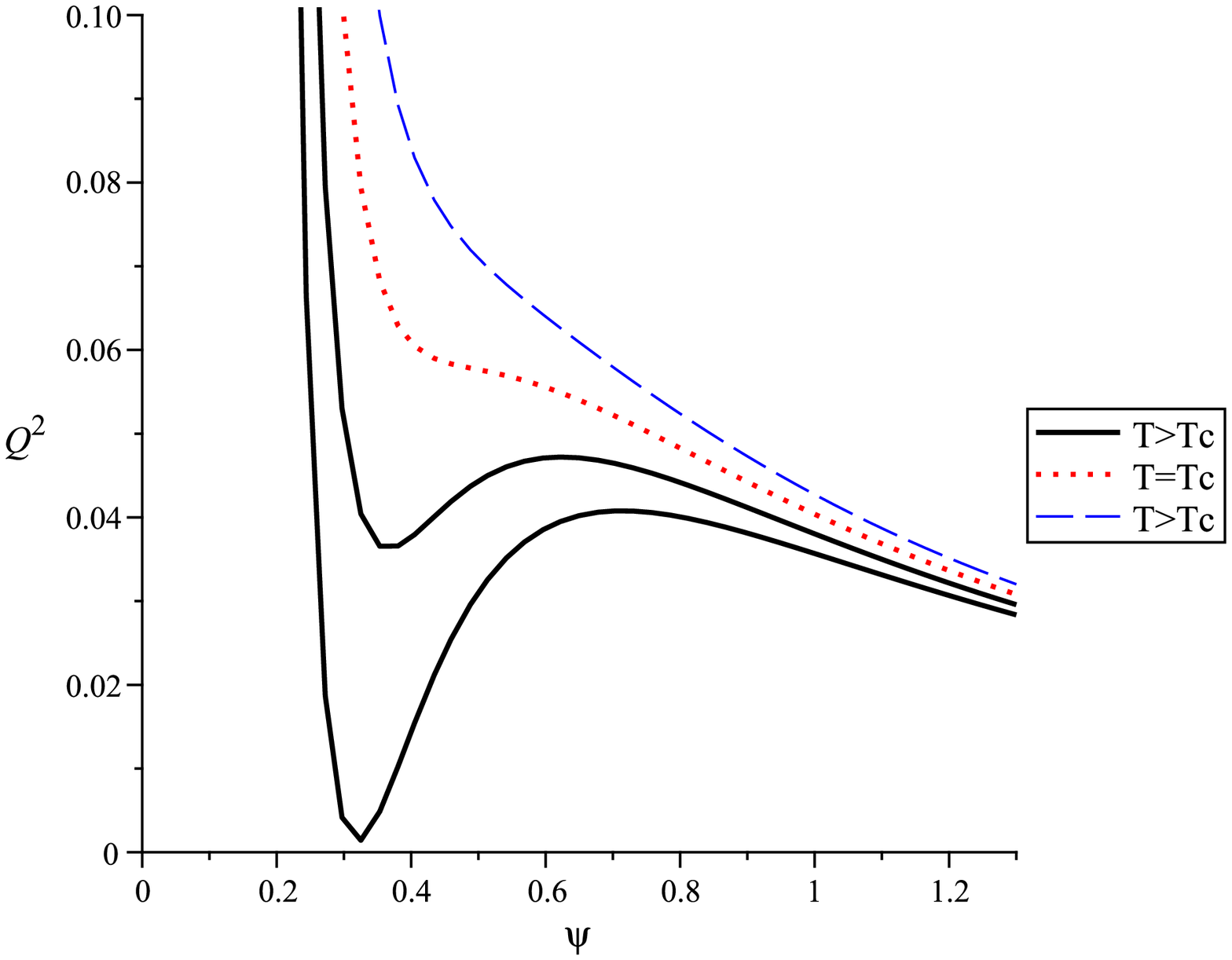}}
  \hspace*{.1cm} \subfigure[$k=1$, $d=6$, $l=1$
]{\includegraphics[scale=0.3]{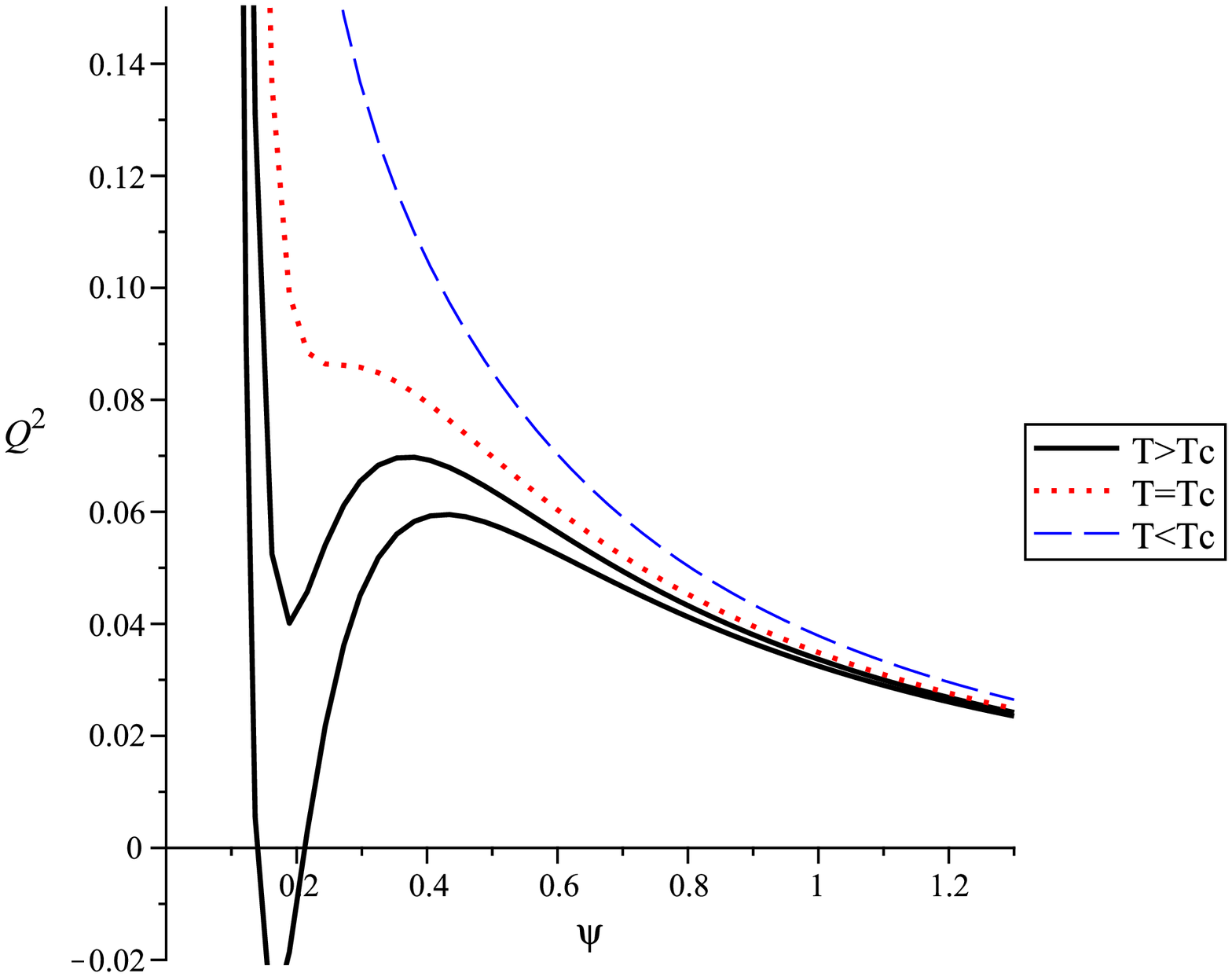}\label{Fig1b}}\caption{$Q^2-\Psi$
diagram of higher dimensional charged AdS black holes.}
    \label{fig1}
  \end{figure}

\begin{figure}
 \centering \subfigure[ $d=5$, $l=1$]
{\includegraphics[scale=0.3]{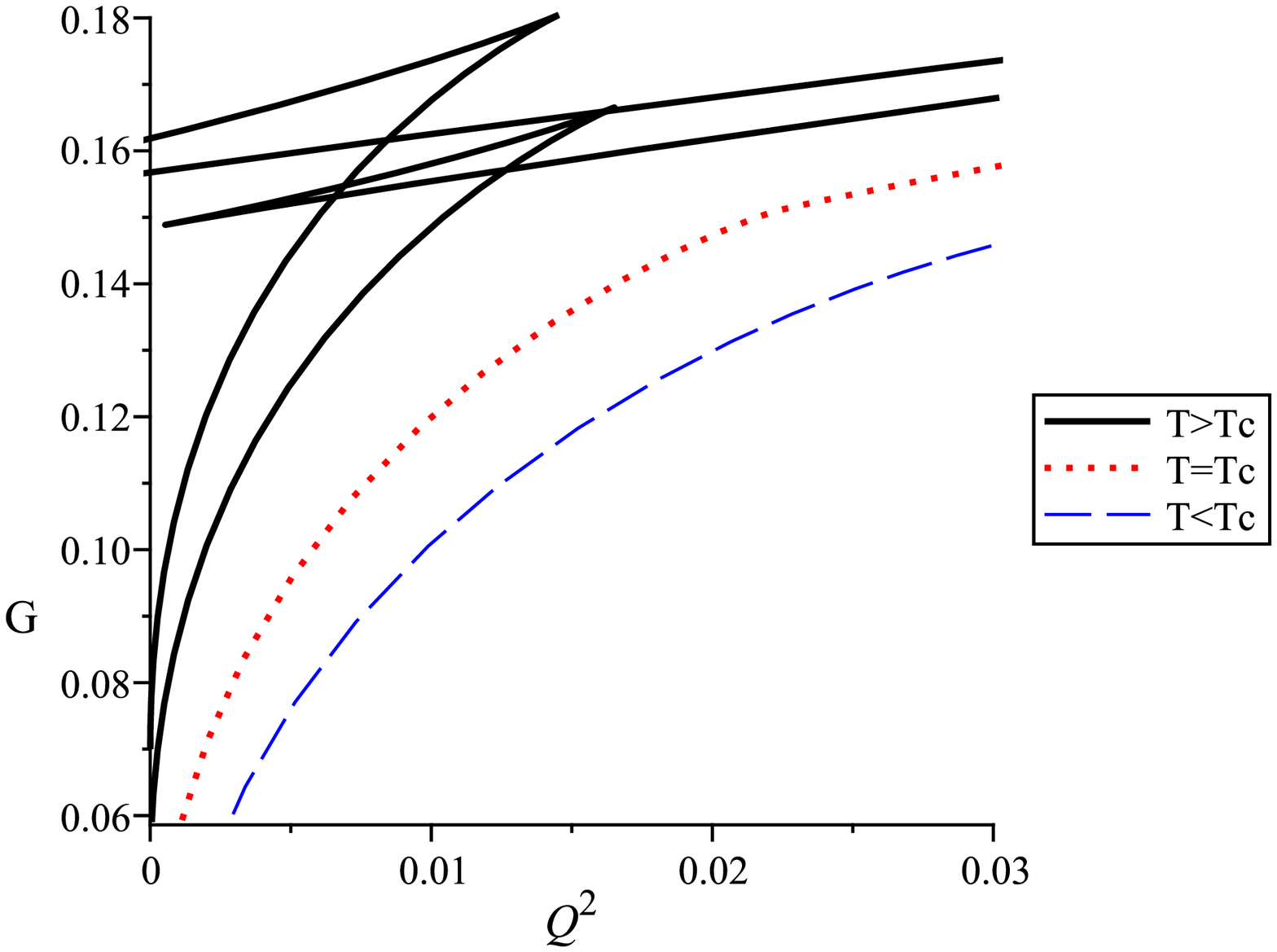}}
  \hspace*{.1cm} \subfigure[ $d=6$, $l=1$
]{\includegraphics[scale=0.3]{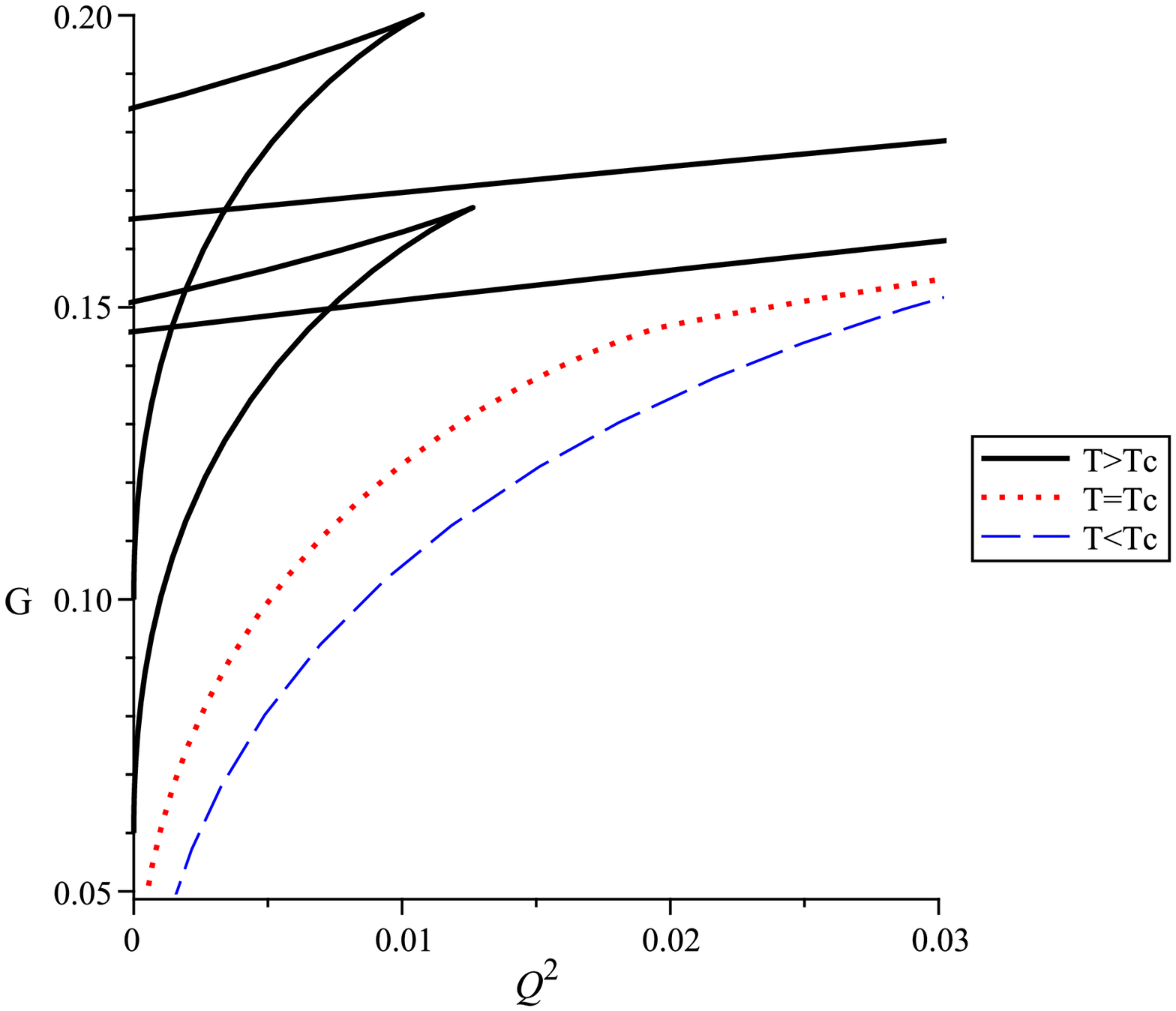}\label{Fig2b}}\caption{$G-Q^2$
diagram of AdS  black holes. The diagrams are shifted for more
clarity.} \label{Fig2}
  \end{figure}

We start by  writing the equation of state in the form
$Q^2(T,\Psi)$ by using Eq.(\ref{T}). For this purpose, we first
write
 \begin{eqnarray}\label{q2}
Q^2(T,r_{+}) =-2T\pi (d-2)
r_{+}^{2d-5}+\frac{(d-1)(d-2)r_{+}^{2d-4}}{2l^2} +\frac{k
(d-3)(d-2)r_{+}^{2d-6}}{2}.
\end{eqnarray}
After replacing $r_+$ from Eq. (\ref{Psi}), one may rewrite the
equation of state as a function of $\Psi$ and $T$,
\begin{eqnarray}\label{eq1}
Q^2(T,\Psi)=-2T\pi (d-2)Y^{2d-5}
\Psi^{\frac{5-2d}{d-3}}+\frac{(d-1)(d-2)}{2l^2}
Y^{2d-4}\Psi^{\frac{4-2d}{d-3}} +\frac{k (d-3)(d-2)}{2
\Psi^2}Y^{2d-6},
\end{eqnarray}
where
\begin{equation}
Y=\left(\frac{\omega_{d-2}}{8\pi (d-3)}\right)^{\frac{1}{d-3}}.
\end{equation}
In order to investigate the critical behavior of the system and
compare with Van der Waals gas, we should plot isotherm diagrams.
The isotherm diagrams $Q^2-\Psi$ given in Fig. \ref{fig1} predict
a first order phase transition in the system which is in complete
analogy with  the Van der Waals liquid-gas system. Note that the
oscillating part $(\frac{\partial Q^2}{\partial \Psi}
>0)$ of the isotherm diagrams show instable regions. The critical point
can be obtained by solving the following equations,
\begin{equation}
\frac{\partial Q^{2}}{\partial \Psi}\Big|_{T_{c}}=0,\quad \frac{\partial ^{2} Q^{2}}{%
\partial \Psi^{2}}\Big|_{T_{c}}=0.  \label{CritEq}
\end{equation}
It is a matter of calculation to show thermodynamic quantities at
the critical point are given by
 \begin{eqnarray}
T_c=\frac{\sqrt{(d-2)(d-1)k}\text{ }(d-3)}{\pi l(2d-5)}.
\end{eqnarray}
 \begin{eqnarray}
\Psi_c=  \frac{\omega_{d-2}
\left[(d-1)(d-2)k\right]^{\frac{(d-3)}{2}}}{8\pi
l^{(d-3)}(d-3)^{(d-2)}}.
\end{eqnarray}
 \begin{eqnarray}
Q^2_c=\frac{l^{(2d-6)}
k^{(d-2)}(d-3)^{(2d-5)}}{2(2d-5)[(d-1)(d-2)]^{(d-3)}}.
\end{eqnarray}
\begin{eqnarray}
\rho_c=T_c\Psi_cQ^2_c=\frac{\omega_{d-2}\text{ }
l^{(d-4)}(d-3)^{(d-2)}\text{ } k^{(d/2)}}{16\pi^2(2d-5)^2
[(d-1)(d-2)]^{\frac{(d-4)}{2}}}.
\end{eqnarray}
In the limiting case where $d=4$ and $k=1$, the above equation
reduces to $\rho_c=1/36\pi$ which is consistent with the result in
\cite{Dehy}. Note that $\rho_c$ is  independent of $l$ only in
four dimension. It is clear from the above equation that the
critical quantities have  reasonable values  only in case $k=1$.
This implies that the system admits a critical behaviour only with
spherical horizon.
The critical behaviour of a system is identified by its partition
function. Indeed, the thermodynamic potential, which is
proportional to the Euclidean action calculating at fixed $Q$ and
$T$, is the Gibbs free energy. To get more information about the
phase transition we calculate the Gibbs free energy $G=M-TS$ as
\cite{Cham}
 \begin{eqnarray}\label{G}
G(Q^2,T)=\frac{\omega_{d-2}}{8\pi (d-3)}\left(k (d-2)(d-3) r^{d-3}- 2\pi (2d-5)T r^{d-2}+\frac{(d-2)^2 r^{d-1}}{l^2}\right),
\end{eqnarray}
where $r_{+}$ is a function of $T$ and $Q^2$ through Eq.
(\ref{q2}). The swallowtail  behavior of the Gibbs free energy in
Fig. \ref{Fig2} represents a first-order phase transition in the
system. A first order phase transition occurs when the Gibbs free
energy is continuous, but its first derivative is discontinuous.
Just like as Van der Waals liquid-gas system.
\subsection{Critical exponents}
Our aim here is to calculate the critical exponents by using the
alternative approach for higher dimensional charged AdS black
holes. The behavior of thermodynamic functions in the vicinity of
the critical point are characterized by the critical exponents. To
find the critical exponent, let us define the
 reduced thermodynamic variables,
\begin{eqnarray}
\Psi_r\equiv \frac{\Psi}{\Psi_c},\quad Q^2_r\equiv
\frac{Q^2}{Q^2_c},\quad T_r\equiv \frac{T}{T_c}.
\end{eqnarray}
Since the critical exponents should be studied near the critical point, we write the reduced variables in the form
\begin{eqnarray}
\Psi_r=1+\psi,\quad Q^2_r=1+\phi,\quad T_r=1+t,
\end{eqnarray}
where $t$, $\psi$ and $\phi$ indicate the deviation from critical
point. One may expand Eq.(\ref{M}) near the critical point and
write
\begin{eqnarray}\label{fi}
\phi=-4(d-2)(d-3)t +4 (d-2)(2d-5)\psi t -\frac{2(d-2)(2d-5)}{3(d-3)^2}\psi^3+o(t \psi^2,\psi^4).
\end{eqnarray}
Using the Maxwell's area law and differentiating Eq.(\ref{fi})
with respect to $\psi$ at a fixed temperature ($t<0$) leads to
\begin{eqnarray}
\phi=-4(d-2)(d-3)t+4 (d-2)(2d-5)\psi_l t -\frac{2(d-2)(2d-5)}{3(d-3)^2}\psi_{l}^3
\nonumber\\=-4(d-2)(d-3)t+4 (d-2)(2d-5)\psi_s t -\frac{2(d-2)(2d-5)}{3(d-3)^2}\psi_{s}^3.
\end{eqnarray}
Indeed
 \begin{eqnarray}
0=\Psi_c \int_{\psi_l}^{ \psi_s} \psi \left(\frac{\partial
Q^2}{\partial \psi}\right)d\psi=\Psi_c \int_{\psi_l}^{ \psi_s}
\psi \left[ 4 (d-2)(2d-5)
t-\frac{6(d-2)(2d-5)}{3(d-3)^2}\psi^2\right] d\psi,
\end{eqnarray}
where $\psi_l$ and $\psi_s$ denote the event horizon of large and
small black hole. The non-trivial solutions of the above equation
are given by
 \begin{eqnarray}\label{sic}
\psi_l=-\psi_s= \sqrt{6(d-3)t}.
\end{eqnarray}
From Eq.(\ref{sic})  we conclude that the order parameter is
$\beta =1/2$. To obtain the critical exponent $\gamma$ the
isotherm compressibility can be calculated as follows
\begin{eqnarray}
\kappa_T=\left(\frac{\partial \Psi}{\partial
Q^2}\right)_T=\frac{\Psi_c }{4 (d-2)(2d-5)Q_c^2 t }\Longrightarrow
\gamma=1.
\end{eqnarray}
In addition it can be easily seen that
\begin{eqnarray}
\phi|_{t=0}= -\frac{2(d-2)(2d-5)}{3(d-3)^2}\psi^3 \Longrightarrow
\delta=3.
\end{eqnarray}
The heat capacity near critical point is $c_\Psi=T\frac{\partial
S}{\partial T}\big|_\Psi=0$. So, the critical exponent $\alpha=0$.
Therefore, we have shown that the critical exponents of the higher
dimensional black holes in the new approach (with fixed $\Lambda$
and  variable $Q^2$) is exactly the same as those presented in
\cite{MannBI} (with $\Lambda$ variable) and coincide with the Van
der Waals fluid system. It is worthwhile to mention that the new
approach is more realistic from physical and mathematical point of
view, because the cosmological constant does not offer a natural
variable.
\section{Thermodynamics of Gauss-Bonnet black holes in AdS space}\label{Structure}
We consider the action of $d$-dimensional
Einstein-Gauss-Bonnet-Maxwell theory in the presence of
cosmological constant $\Lambda$ which can be written
\begin{eqnarray}
S =\frac{1}{16\pi}\int
d^dx\sqrt{-g}[{R-2\Lambda-\alpha_{GB}\left(R_{\mu\nu\gamma\delta}R^{\mu\nu\gamma\delta}-4R_{\mu\nu}
R^{\mu\nu}+R^2\right) -4\pi F_{\mu\nu} F^{\mu\nu}}],
\label{Stress}
\end{eqnarray}
where $ \alpha_{GB}$ is the GB coefficient with dimension of
[length]$^2$ which is proportional to inverse string tension with
positive coefficient \cite{Bou}. The metric function $f(r)$ of
charged GB black holes in AdS space is given by \cite{Cai2}
\begin{eqnarray}
f(r)= k+
\frac{r^2}{2\tilde{\alpha}}\Bigg[1-\sqrt{1-\frac{8\tilde{\alpha} Q^2}{(d-2)(d-3)r^{2d-4}}
    +\frac{64\pi\tilde{\alpha} M}{(d-2)\omega_{d-2} r^{d-1}}-\frac{64\pi\tilde{\alpha} P}{(d-2)(d-1)}} \ \ \Bigg].
\end{eqnarray}
where $\tilde{\alpha}=(d-3)(d-4)\alpha_{G\beta}$ and $k$
represents the topology of the horizon, and we have replaced
$\Lambda$ with $P$ by using Eq. (\ref{pV}). The constant $M$ is
the mass, while $Q$ is related to the charge of the black hole.
The position of the black hole event horizon is determined as a
larger root of $f(r_+)=0$ and hence the mass of black hole which
is equivalent with enthalpy is calculated as \cite{Cai2}
\begin{eqnarray}\label{mm}
H\equiv M =\frac{(d-2)\omega_{d-2}r_{+}^{(d-3)}}{16
\pi}\left(k+\frac{k^2\tilde{\alpha}}{r_{+}^2}+\frac{16\pi P
r_{+}^2}{(d-1)(d-2)}+\frac{2Q^2r_{+}^{6-2d}}{(d-2)(d-3)}\right),
\end{eqnarray}
where $P$ is pressure and defined in (\ref{pV}). The temperature
and entropy of the black hole  can be given by
\begin{eqnarray}
T=\frac{f'( r_{+})}{4 \pi} =\frac{\frac{16\pi P
r_{+}^4}{(d-2)}+(d-3)k
r_{+}^2+(d-5)k^2\tilde{\alpha}-\frac{2Q^2}{(d-2)r_{+}^{(2d-8)}
}}{4\pi  r_{+}( r_{+}^2+2k\tilde{\alpha})},\label{Temp}
\end{eqnarray}
and
 \begin{eqnarray}
S=\int_{0}^{ r_{+}} T^{-1} \left(\frac{\partial M}{\partial
r}\right)_{Q,P}dr=\frac{\omega_{d-2}r_{+}^{(d-2)}}{4}\left[1+\frac{2(d-2)\tilde{\alpha} k}{(d-4)
r_{+}^2}\right].
\end{eqnarray}
We can also calculate the thermodynamic volume $V$  and the
electric potential $\Phi$ as
\begin{eqnarray}
V= \left (\frac{\partial M}{\partial
P}\right)_{Q^2,S}=\frac{\omega_{d-2} r_{+}^{(d-1)}}{d-1},
\end{eqnarray}
\begin{eqnarray}
\Phi=\frac{Q\omega_{d-2}}{4\pi(d-3)r_{+}^{d-3}}.
\end{eqnarray}
These quantities satisfy the first law of black holes
thermodynamics \cite{Cai2}
\begin{eqnarray}
dM=TdS+\Phi dQ+VdP+\Omega d\tilde{\alpha},
\end{eqnarray}
where
\begin{eqnarray}
\Omega= \left(\frac{\partial M}{\partial
\tilde{\alpha}}\right)_{S,Q,P},
\end{eqnarray}
is the quantity conjugate to the GB coefficient $\tilde{\alpha}$.
Since $\tilde{\alpha}$ is a dimensionful parameter, the
corresponding term will inevitably appear in the Smarr formula
\cite{Cai2}
\begin{equation}
M=\frac{d-2}{d-3}T S+\Phi Q-\frac{2}{d-3}V P + \frac{2}{d-3} \Omega \tilde{\alpha}.
\end{equation}
As mentioned before, we want to study the phase structure with a
different point of view which $Q^2$ plays the role of thermodynamic
variable. We also, define  a new
variable in an our alternative approach, namely $\Psi$ which is
\begin{eqnarray}
\Psi=\left(\frac{\partial M}{\partial
Q^2}\right)_{P,S}=\frac{\omega_{d-2}r_{+}^{(3-d)} }{8\pi
(d-3)}.\label{psi}
\end{eqnarray}
The new variable $\Psi$, pressure $P$  and temperature $T$ are
intensive parameters conjugate to $Q^2$, $V$ and $S$ respectively.
We also replace the term $\Phi dQ$ by $\Psi dQ^2$ in the first law
of thermodynamics. By using (\ref{psi}), one may easily show that
the Smarr formula and the first law of thermodynamics are now
given by
\begin{equation}
M=\frac{d-2}{d-3}T S+\Psi Q^2 -\frac{2}{d-3}V P + \frac{2}{d-3} \Omega \tilde{\alpha}
\end{equation}
and
\begin{eqnarray}
dM=TdS+\Psi dQ ^2+VdP+\Omega d\tilde{\alpha}.
\end{eqnarray}

\section{Critical behavior of GB black holes in Five dimension}\label{d5}
In order to study the critical behaviour of GB black hole using
the alternative approach, we first consider the case $d=5$
dimension. In this case, the equation of state is simple enough so
that we can solve the equations and obtain critical quantities
exactly. Since in this approach the cosmological constant
(pressure) is regarded a constant quantity, so we consider the
case with $P=0$ and $P\neq0$, separately.
\subsection{Critical behavior with $P=0$}
Setting $P=0$ and $d=5$, the mass of the GB black hole given in
Eq.(\ref{mm}) can be written
 \begin{eqnarray}
M= \frac{3 \tilde{\alpha}  k^2 \omega_{3}}{16 \pi }+\frac{3 k
r_{+}^2
   \omega_{3}}{16 \pi }+\frac{{Q^2} \omega_{3}}{64
   \pi  r_{+}^2}.
\end{eqnarray}
Combining the Hawking temperature given in Eq.(\ref{Temp}) with
the above condition, the equation of state of the black hole can
be written as
 \begin{eqnarray}\label{qqk}
Q^2(\Psi, T)=\frac{3 k {\omega_{3}}^2}{256 \pi ^2 \Psi ^2}-\frac{3
   \tilde{\alpha}  k T {\omega_{3}}^{3/2}}{16 \sqrt{\pi }
    \Psi
   ^{3/2}}-\frac{3 T {\omega_{3}}^{5/2}}{512 \pi ^{3/2}
    \Psi
   ^{5/2}}.
\end{eqnarray}
Critical points occur at stationary points of inflection in
$Q^2-\Psi$ diagram, where
\begin{equation}
\frac{\partial Q^{2}}{\partial \Psi}\Big|_{T_{c}}=0,\quad \frac{\partial ^{2} Q^{2}}{%
\partial \Psi^{2}}\Big|_{T_{c}}=0. \label{crit eq}
\end{equation}
In case of spherical horizon where $k=1$ and $\omega_{3}=2 \pi^2$,
one may obtain the critical quantities of GB black hole as
 \begin{eqnarray}
T_c=\frac{1}{\pi \sqrt{30 \tilde{\alpha}}},\quad \Psi_c= \frac{5
\pi}{48 \tilde{\alpha}},\quad Q^2_c=-\frac{36
\tilde{\alpha}^2}{125}.
\end{eqnarray}
We see that in this case  $Q^2_c$ is negative, which is physically
not acceptable. Therefore, we conclude that in this case there
does not exit any phase transition, and therefore $Q^2-\Psi$
diagram, has no similarity with isotherm diagrams of Van der Waals
system. Let us check whether or not there is phase transition or
critical behavior for other topology of horizon namely flat
$(k=0)$ or hyperbolic $(k=-1)$ cases.

In hyperbolic $(k=-1)$ case, the equation of state reduces to
 \begin{eqnarray}\label{qqd5p}
Q^2(\Psi, T)=-\frac{3 {\omega_{3}}^2}{256 \pi ^2 \Psi ^2}+\frac{3
   \tilde{\alpha}  T {\omega_{3}}^{3/2}}{16 \sqrt{\pi }
    \Psi
   ^{3/2}}-\frac{3 T {\omega_{3}}^{5/2}}{512 \pi ^{3/2}
    \Psi
   ^{5/2}}.
\end{eqnarray}
In this equation the value of positive term is smaller than the
other two terms with negative values because $\tilde{\alpha}$
always has positive small value $(0<\tilde{\alpha} <1)$. Therefore
$Q^2$ is a monotonic function of $\Psi$ and there is no critical
point and phase transition. For the flat horizon, the equation of
state  reads
 \begin{eqnarray}
Q^2(\Psi, T)= -\frac{3 \pi ^{7/2} T}{64 \sqrt{2} \Psi ^{5/2}}.
\end{eqnarray}
It is clear that this equation is monotonic  function which can
not cause any  phase transition. One may conclude that the
existence of $\Lambda$ (pressure) is essential for the critical
behavior in this new standpoint. Similary, the existence of $Q$ is
necessary when we treat $\Lambda$  as a thermodynamic variable
\cite{Mann1}. It may show a meaningful symmetry between new
approach with fixed $\Lambda$ and old one with fixed $Q$.
\subsection{Critical behavior with $P\neq0$ }
\begin{figure}[tbp]
 \centering
{\includegraphics[scale=0.4]{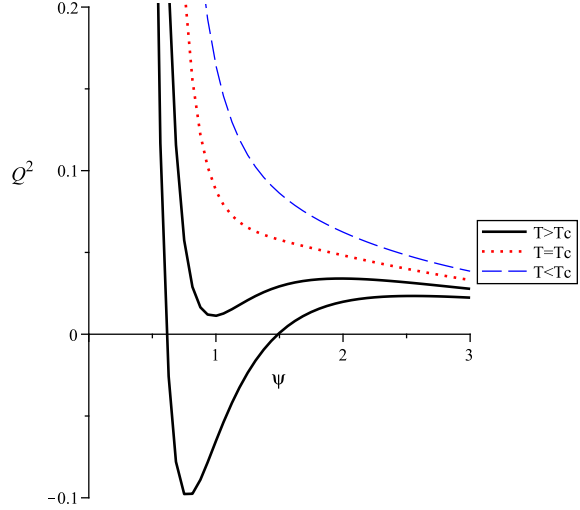} } \hspace*{.1cm}
\caption{$Q^2-\Psi$ diagram of GB black holes for $k=1$, $d=5$,
$l=1$, $\tilde{\alpha}=0.01$.} \label{Fig3}
\end{figure}

\begin{figure}
\centering {\includegraphics[scale=0.4]{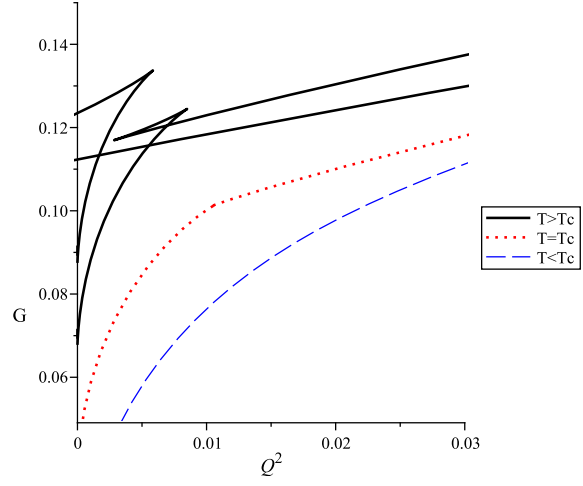}} \hspace*{.1cm}
\caption{$G-Q^2$ diagram of GB black holes for $d=5$, $l=1$,
$\tilde{\alpha}=0.01$.The diagrams are shifted for more clarity.}
\label{Fig4}
\end{figure}

Following the approach taken previously, one may find no critical
behavior in the cases with for $(k=0,-1 )$. Thus, we just focus on
black holes with spherical topology ($k=1$). In this case
Eq.(\ref{mm}) with new conditions, takes the form
\begin{eqnarray}
M=\frac{3}{8} \pi  \tilde{\alpha}  k^2+\frac{3}{8} \pi  k
r_{+}^2+\frac{1}{2} \pi ^2 P r_{+}^4+\frac{\pi  Q^2}{8 r_{+}^2},
\end{eqnarray}
where the relation between $r_{+}$ and new parameter $\Psi$ is
$r_{+}=\frac{\sqrt{\pi }}{2 \sqrt{2 \Psi }}$. Using the Hawking
temperature in Eq.(\ref{Temp}), the equation of state may be
obtained as
\begin{eqnarray}\label{QQ}
Q^2(\Psi, T)=\frac{\pi ^2 }{128 \Psi ^3}\left(2 \pi ^2 P-48
\sqrt{2\pi}  \Psi ^{3/2} \tilde{\alpha} T-3 \sqrt{2} \pi ^{3/2}
\Psi^{1/2 }T+6 \Psi \right)\label{eq state d=5}
\end{eqnarray}
As isotherm diagram shows in Fig.\ref{Fig3}, for  constant
pressure and $T=T_c$, there is an inflection point in $Q^2-\Psi$
diagrams which is the critical point where the second order phase
transition occurs. The critical values reads
\begin{eqnarray}
T_c=\frac{\left[(3 \Gamma+25) l^2-162 \tilde{\alpha} \right]
   }{100 \sqrt{3}
   \pi  \sqrt{\tilde{\alpha} } l^2} \text{ }\sqrt{-\Gamma-\frac{54 \tilde{\alpha} }{l^2}+5},
\end{eqnarray}
\begin{eqnarray}
\Psi_c=-\frac{\pi  \left[54 \tilde{\alpha} +(\Gamma-5) l^2\right]}{96
   \tilde{\alpha}  l^2},
\end{eqnarray}
\begin{eqnarray}
Q^2_c=-\frac{144\tilde{\alpha} ^2 l^4 \left[126 \tilde{\alpha} +(\Gamma-5)
   l^2\right]}{5 \left[54 \tilde{\alpha} +(\Gamma-5)
   l^2\right]^3},
\end{eqnarray}
where
\begin{equation}
\Gamma=\sqrt{\frac{36 \tilde{\alpha}  \left(81 \tilde{\alpha} -25
   l^2\right)}{l^4}+25}.
\end{equation}
We can also find
\begin{eqnarray}
\rho_c=T_c \Psi_c Q^2_c=\frac{\sqrt{3 \tilde{\alpha}}
\left[(3 \Gamma+25) l^2-162 \tilde{\alpha} \right] \left[126
\tilde{\alpha} +(\Gamma-5) l^2\right] }{1000 \left[54 \tilde{\alpha}
+(\Gamma-5) l^2\right]^2} \sqrt{-\Gamma-\frac{54 \tilde{\alpha} }{l^2}+5}.
\end{eqnarray}
It can be realized that the above equations admit a phase
transition with acceptable critical quantities provided the
following two constrains are satisfied
\begin{equation}\label{lag}
\left\{
  \begin{array}{ll}
    $$\frac{36 \tilde{\alpha}  \left(81 \tilde{\alpha} -25
       l^2\right)}{l^4}+25 \geq 0,\quad \quad\quad $$  &  \\
    &\\
$$ -\Gamma-\frac{54 \tilde{\alpha} }{l^2}+5\geq 0,\quad~ \quad\quad \quad$$
  \end{array}
\right.
\end{equation}
As one can see these conditions depend on the  amount of $l$ and
$\tilde{\alpha}$. However, they will be satisfied automatically
for small $\tilde{\alpha}$ as one may see in the next section.  In
the limit of small $\tilde{\alpha}$, the series expansions of
critical quantities are
\begin{eqnarray}
T_c=\frac{4 \sqrt{3}}{5 \pi l}-\frac{72 \sqrt{3} \tilde{\alpha}
}{25 \pi l^3},\quad \Psi_c=\frac{27 \pi  \tilde{\alpha} }{5
l^4}+\frac{3 \pi }{8 l^2} ,\quad Q^2_c=\frac{l^4}{45}-\frac{32
\tilde{\alpha}  l^2}{25},
\end{eqnarray}
which reduces to the following equation for small $\tilde{\alpha}$
\begin{eqnarray}
\rho_c=T_c \Psi_c Q^2_c=\frac{l}{50 \sqrt{3} }-\frac{39}{125}
\left(\frac{ \sqrt{3}}{l}\right) \tilde{\alpha}
+O(\tilde{\alpha}^2).
\end{eqnarray}
In the absence  GB correction terms ($\tilde{\alpha}=0$) all
critical values reduce to the results of section \ref{Field}. The
critical behavior of a thermodynamic system can be characterized
by the Gibbs free energy, which in our case it can be written as
\begin{eqnarray}
G(Q^2,T)=\frac{3 \pi  \tilde{\alpha} }{8}+\frac{9 \pi  r_{+}^4}{16
   l^2}-\frac{1}{16} \pi ^2 r_{+} T \left(54 \tilde{\alpha} +11
   r_{+}^2\right)+\frac{15 \pi  r_{+}^2}{32}.
\end{eqnarray}
The behavior of the Gibbs free energy is depicted in
Fig.\ref{Fig4} in terms of $Q^2$ for various temperature.
Evidently, for $T>T_c$  the Gibbs free energy develops a
swallowtail shape which shows first order phase transition. The
critical behavior of GB black hole ($k=1$) can be characterized by
the critical exponent. In order to examine the critical exponents
we introduce the following reduced thermodynamic variables
\begin{eqnarray}
\Psi_r\equiv \frac{\Psi}{\Psi_c}=1+\psi,\quad Q^2_r\equiv
\frac{Q^2}{Q^2_c}=1+\phi,\quad T_r\equiv \frac{T}{T_c}=1+t.
\end{eqnarray}
The Taylor expansion of Eq.(\ref{QQ}) is
\begin{eqnarray}\label{abc}
\phi=A t +B\psi t - C\psi^3+o(t \psi^2 ,\psi^4),
\end{eqnarray}
where $A=\left(-24-\frac{576\tilde{\alpha}}{l^2}\right)$,
$B=\left(60+\frac{1296\tilde{\alpha}}{l^2}\right)$ and
$C=\left(\frac{5}{2}+\frac{18\tilde{\alpha}}{l^2}\right)$.
Applying Maxwell's equal area law and considering the fact that
during the phase transition the charge remains constant we have
\begin{eqnarray}\label{abcabc}
A t +B\psi_l t - C\psi_l^{3}=A t +B\psi_s t - C\psi_s^3.
\end{eqnarray}
 \begin{eqnarray}\label{max}
0=\Psi_c \int_{\psi_l}^{ \psi_s} \psi \left(\frac{\partial
Q^2}{\partial \psi}\right )d\psi=\Psi_c \int_{\psi_l}^{ \psi_s}
\psi (B t-3 C \psi^2)d\psi.
\end{eqnarray}
The nontrivial solution for Eqs.(\ref{abcabc}) and (\ref{max})
reads
 \begin{eqnarray}
\psi_l=- \psi_s \longrightarrow |\psi_l-\psi_s|= 2
\sqrt{\frac{-B}{C}}t^{\frac{1}{2}}.
\end{eqnarray}
We conclude that the order parameter $\beta$ which is appropriate
with the power of $t$ is $(\beta=\frac{1}{2})$. The next critical
exponent is $\gamma$ which can be obtained by the following
relation
\begin{equation}
\chi _{T}=\frac{\partial \Psi}{\partial Q^2}\Big|_{T}\propto \frac{
\Psi_{c}}{B Q^2_{c}}\frac{1}{t}\quad \Longrightarrow \quad \gamma=1
\end{equation}
The shape of the critical isotherm at $t=0$ is given by
Eq.(\ref{abc}). We find
\begin{equation}
\phi|_{t=0}=-C\psi ^{3}\quad \Longrightarrow \quad \delta =3.
\end{equation}
Finally, the heat capacity near the critical point at fixed $\Psi$ reads
 \begin{equation}
c_\Psi=T\frac{\partial S}{\partial T}\big|_\Psi=0
\end{equation}
Since the entropy is independent of $T$, the critical exponent
$\alpha=0$. Therefore, we have obtained all critical exponents in
$Q^2-\Psi$ plans for GB black holes in five dimensions with
spherical horizon. We have treated the charge of the black hole as
a thermodynamic variable and kept the pressure as a fixed
parameter. We have also confirmed that these critical exponents
are similar to those of Van der Waals liquid-gas system.
\section{Critical behavior of Gb black holes in arbitrary dimensions}\label{arbitrary d}
\begin{figure}
 \centering \subfigure[ $k=1$, $d=6$, $l=1$, $\tilde{\alpha}=0.01$]
{\includegraphics[scale=0.3]{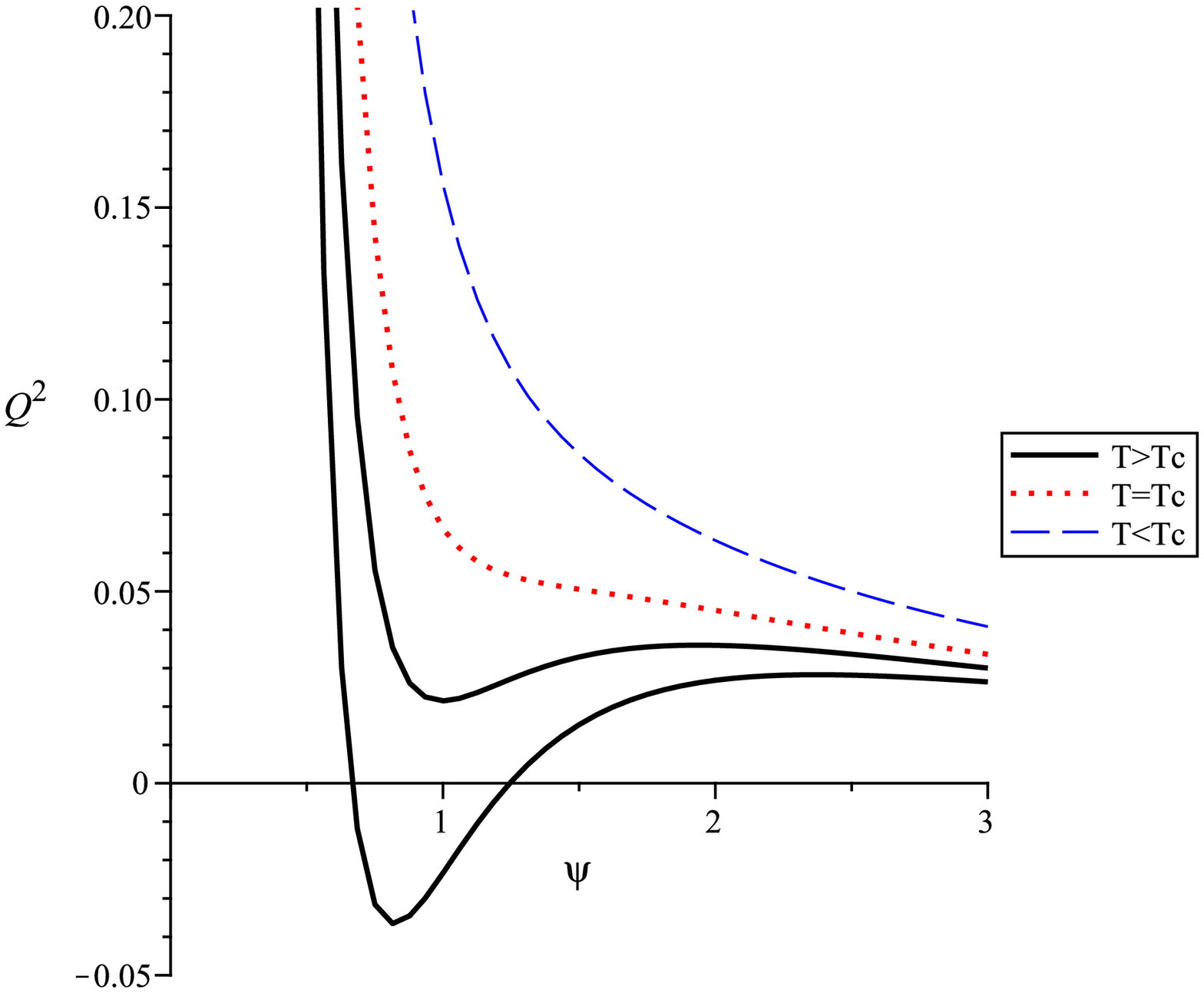}\label{Fig5a}}
  \hspace*{.1cm} \subfigure[$k=1$, $d=7$, $l=1$,$\tilde{\alpha}=0.01$]
{\includegraphics[scale=0.3]{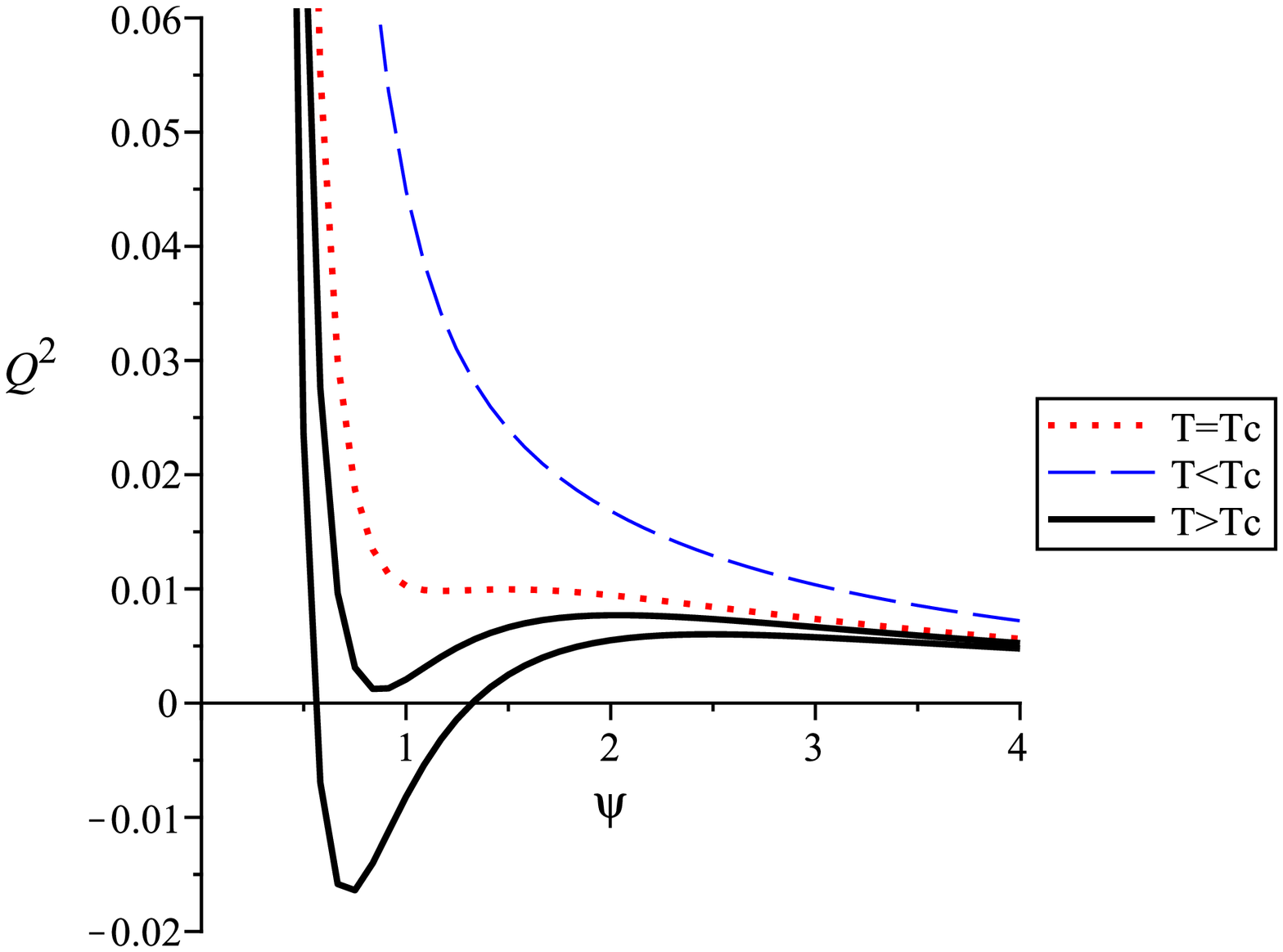}\label{Fig5b}}\caption{$Q^2-\Psi$ diagram of Guass-Bonnet black holes in higher dimensions.}
    \label{Fig5}
  \end{figure}

\begin{figure}
 \centering \subfigure[$d=6$, $l=1$, $\tilde{\alpha}=0.01$]
{\includegraphics[scale=0.3]{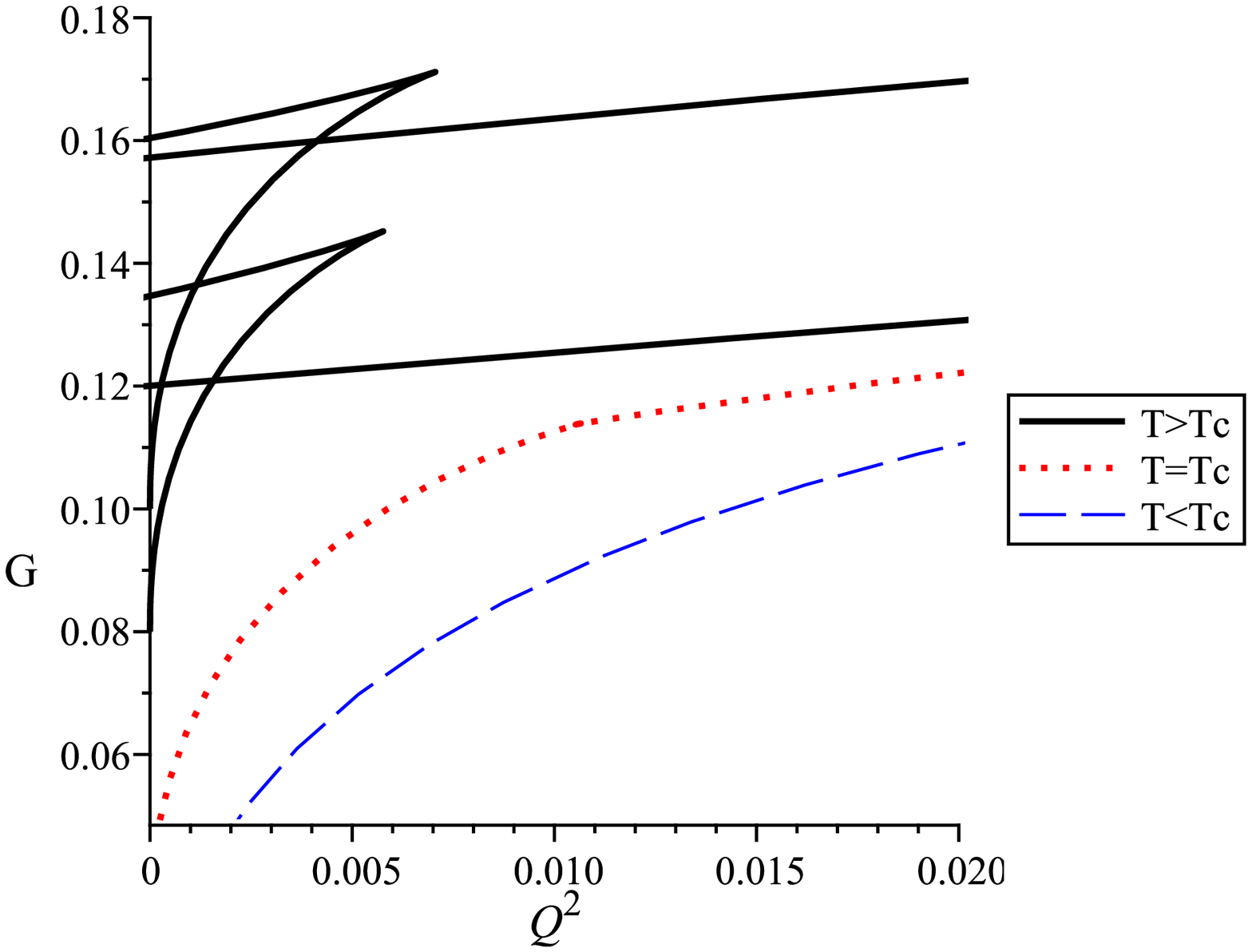}\label{Fig6a}}
  \hspace*{.1cm} \subfigure[$d=7$, $l=1$, $\tilde{\alpha}=0.01$
]{\includegraphics[scale=0.3]{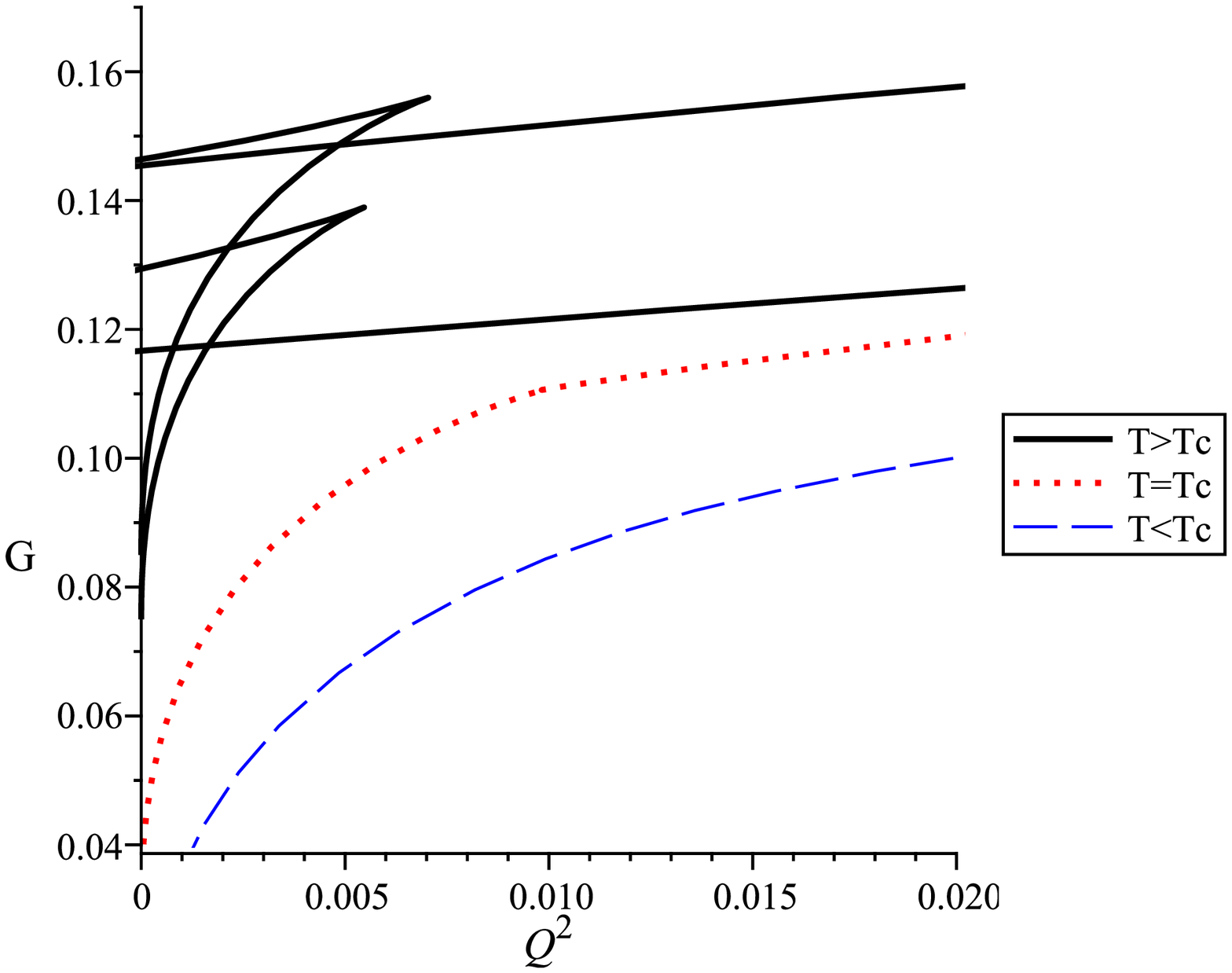}\label{Fig6b}}\caption{$G-Q^2$ diagram of Guass-Bonnet black holes which shifted for more clarity.}
    \label{Fig6}
  \end{figure}

In this section we are going to extend our investigation on the
critical behavior of GB black hole to all higher dimensions. Our
approach for calculating the critical quantities and critical
exponents in $d>5$ is exactly the same as in five dimensions.
Therefore, for the economic reasons we do not repeat the
calculations and only give the results. Using the Hawking
temperature (\ref{Temp}), we can write the equation of state in
$d$-dimensions as
\begin{eqnarray}
Q^2(\Psi,T)&=& \frac{1}{2} Y^{2d-8} \Psi^{\frac{2d-8}{3-d}}
\left((d-5)(d-2)k^2 \tilde{\alpha}+(d-2) k
\Psi^{\frac{1}{3-d}}\Bigg(-8\pi T \tilde{\alpha}+ Y(d-3)
\Psi^{\frac{1}{3-d}}\right)\nonumber\\&& +4 Y^3 \pi
\Psi^{\frac{3}{3-d}}\left(-(d-2) T +4 Y
P\Psi^{\frac{1}{3-d}}\right )\Bigg).
\end{eqnarray}
The behavior of the isotherm diagrams in $d=6,7$ are shown in
Fig.\ref{Fig5} which show the same behavior as Van der Waals
system and predict a first order phase transition in the system.
The equation of state is complicated and it is not easy to obtain
the critical point analytically, however, we can still calculate
the critical quantities for small $\tilde{\alpha}$. As we have
explained in the previous section, we do not expect to see
critical behavior in the absence of cosmological constant ($P=0$),
or in case with $k=0, -1$. Thus, we consider the spherical horizon
in the presence of $\Lambda$. Using Eq. (\ref{crit eq}) in case
$k=1$, one can find
 \begin{eqnarray}
T_c&=&\frac{d^3-6 d^2+11 d-6}{\pi  \sqrt{(d-1)(d-2)}  (2 d-5)
l}+\frac{\tilde{\alpha} (d-2)^{3/2} (d-1)^{3/2} \left(-6 d^3+53
d^2-155 d+152\right)}{2 \pi  (5-2 d)^2 (d-3)^3
l^3}+O(\tilde{\alpha}^2),\\
\Psi_c &=&\frac{(d-3)^{2-d} (d-2)^{\frac{d-3}{2}}
(d-1)^{\frac{d-3}{2}} \omega_{d-2} l^{3-d}}{8 \pi } \nonumber\\&&
-\frac{(d-3)^{(-1-d)} (d-2)^{\frac{d-1}{2}}\label{eq d}
(d-1)^{\frac{d-1}{2}}}{16 \pi (2d-5)}
\left[204+d(-225+(83-10d)d)\right] l^{1-d} \omega_{d-2}
\tilde{\alpha}+O(\tilde{\alpha}^2),\\
Q^2_c &=&\frac{(d-3)^{2 d-5} (d-2)^{3-d} (d-1)^{3-d} l^{2 d-6}}{2 (2
   d-5)} \nonumber\\&&
-\frac{\tilde{\alpha}  (d-3)^{2 d-8} (d-2)^{5-d} (d-1)^{4-d}
   \left(10 d^2-51 d+69\right) l^{2 d-8}}{2 (5-2 d)^2}+O(\tilde{\alpha}^2),
\end{eqnarray}
which leads to
\begin{eqnarray}
\rho_c&=&Q^2_c Tc \Psi_c= \frac{(d-3)^{d-2} (d-2)^{2-\frac{d}{2}} (d-1)^{2-\frac{d}{2}}
   \omega_{d-2} l^{d-4}}{16 \pi ^2 (2 d-5)^2}- \\ && \nonumber\frac{\tilde{\alpha}
   \left((d-3)^{d-6} (d-2)^{3-\frac{d}{2}}
   (d-1)^{3-\frac{d}{2}} \left(10 d^4-83 d^3+241 d^2-268
   d+64\right) \omega_{d-2} l^{d-6}\right)}{32 \pi ^2 (2
   d-5)^3}+O\left(\tilde{\alpha} ^2\right)
\end{eqnarray}
It is worthwhile to note that all above equations reduce to the
results in section \ref{Field} when $\tilde{\alpha}$. The Gibbs
free energy of the GB black hole $G=M-T S$ can be calculated as
\begin{eqnarray}
G(Q^2,T)&=&\frac{\omega_{d-2} }{64\pi (d-3)}\Bigg(\frac{(d-2) (5 d-13) r^{d-1}}{ l^2}+r^{d-3} \left[5 d (d-4 \pi  r T-5)+56
   \pi  r T+30 \right]   \notag \\
    &&
   +\frac{\tilde{\alpha} (d-2) r^{d-5} \left(5
   d^2+8 \pi  (16-5 d) r T-37 d+68\right)}{
   (d-4) }\Bigg) + O\left(\tilde{\alpha} ^2\right)
   \end{eqnarray}
Fig. \ref{Fig6} shows the behaviour of Gibbs free energy versus
$Q^2$ which shows that the phase transition can occur when the
temperature is more than $T_c$. From these figures we find that GB
black holes admit a first order  phase transition in higher
dimensions. Following the method of the previous section, we can
calculate the critical exponents of GB black holes in higher
dimensions. The result is
\begin{eqnarray}
\alpha=0,\quad \beta=1/2, \quad \gamma=1,\quad \delta=3.
\end{eqnarray}

\begin{figure}
 \centering \subfigure[ $d=5$, $m=1$]
 {\includegraphics[scale=0.3]{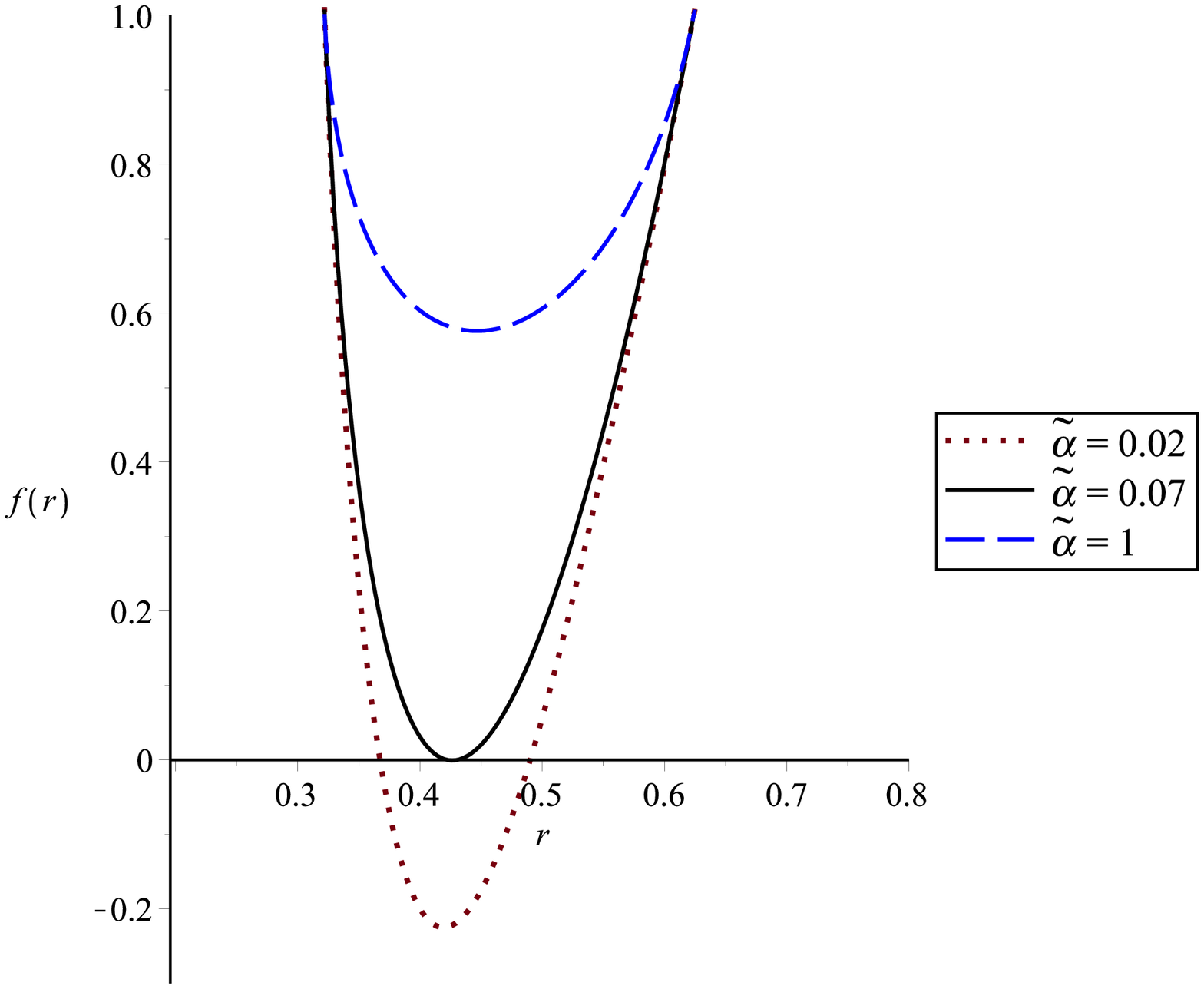}\label{Fig7a}}
  \hspace*{.1cm} \subfigure[$d=6$, $m=1.2$
]{\includegraphics[scale=0.3]{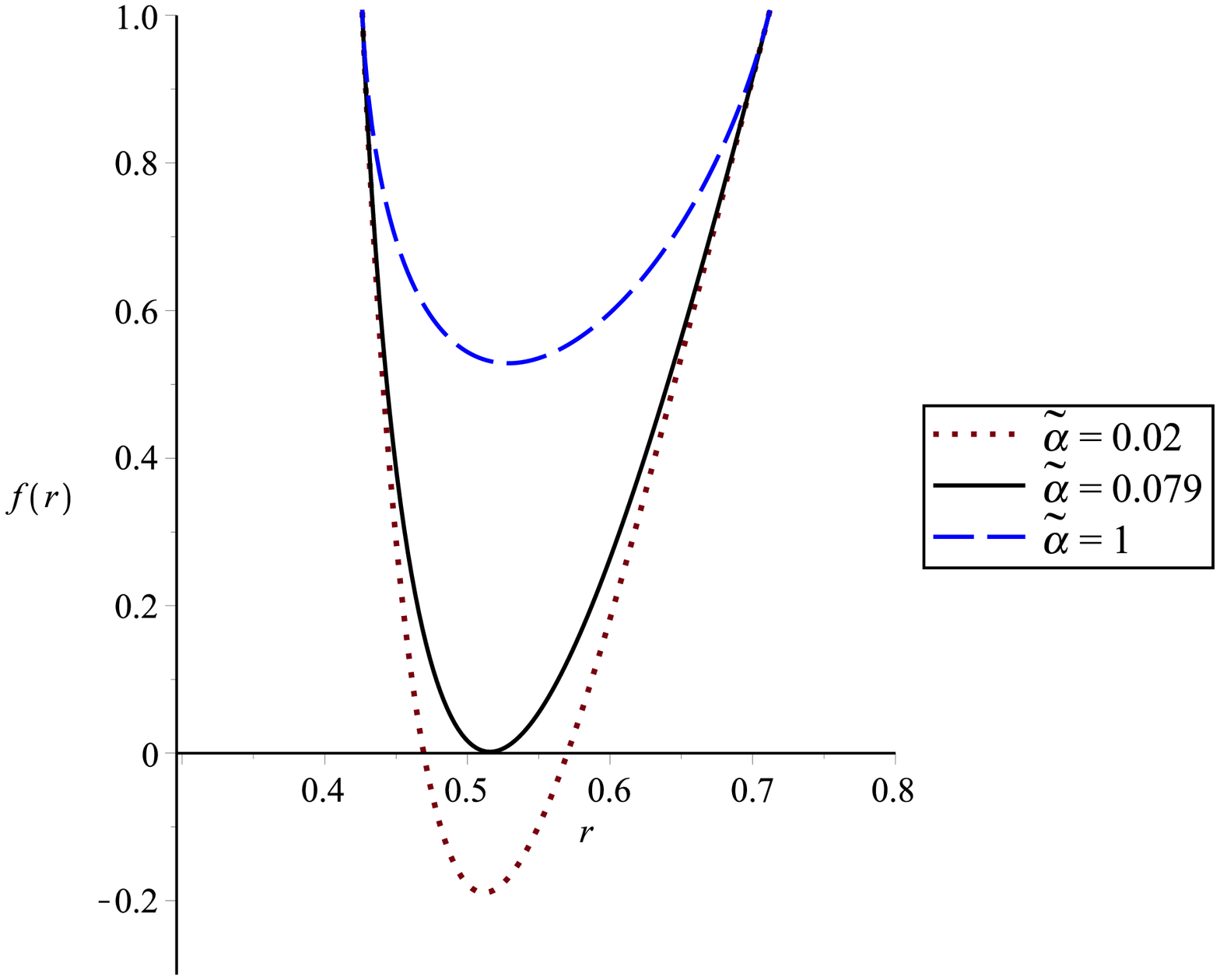}\label{Fig7.b}}\caption{$f(r)$
versus $r$ for different values of  $\tilde{\alpha}$ with
$k=l=1$.}
\label{Fig7}
   \end{figure}

A close look at the behavior of $f(r)$ in Figs.\ref{Fig7}, shows the existence of zero, one or two roots for the
metric function depending on the value of $\tilde{\alpha}$. It is
worthwhile to note that the event horizon disappears with
increasing $\tilde{\alpha}$. So we have a naked singularity when
$\tilde{\alpha}$ is larger than specific value. In other words, we
did not see any critical behavior for large value of
$\tilde{\alpha}$ because there is no event horizon in this range
of $\tilde{\alpha}$.

\begin{figure}
 \centering \subfigure[$T_c$ versus $\tilde{\alpha}$ ]
{\includegraphics[scale=0.3]{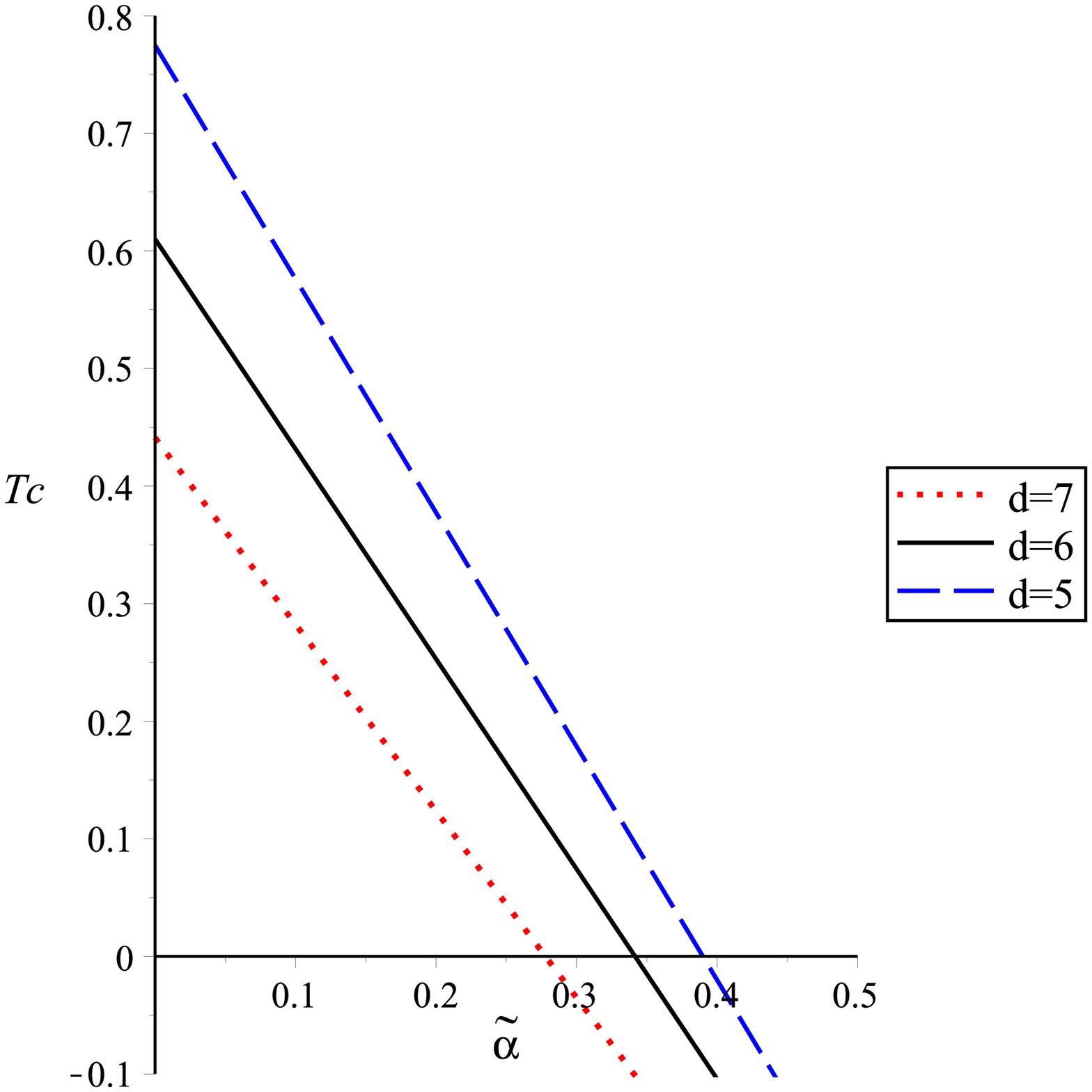}\label{Fig9a}}
  \hspace*{.1cm} \subfigure[$Q^2_c$ versus $\tilde{\alpha}$
]{\includegraphics[scale=0.3]{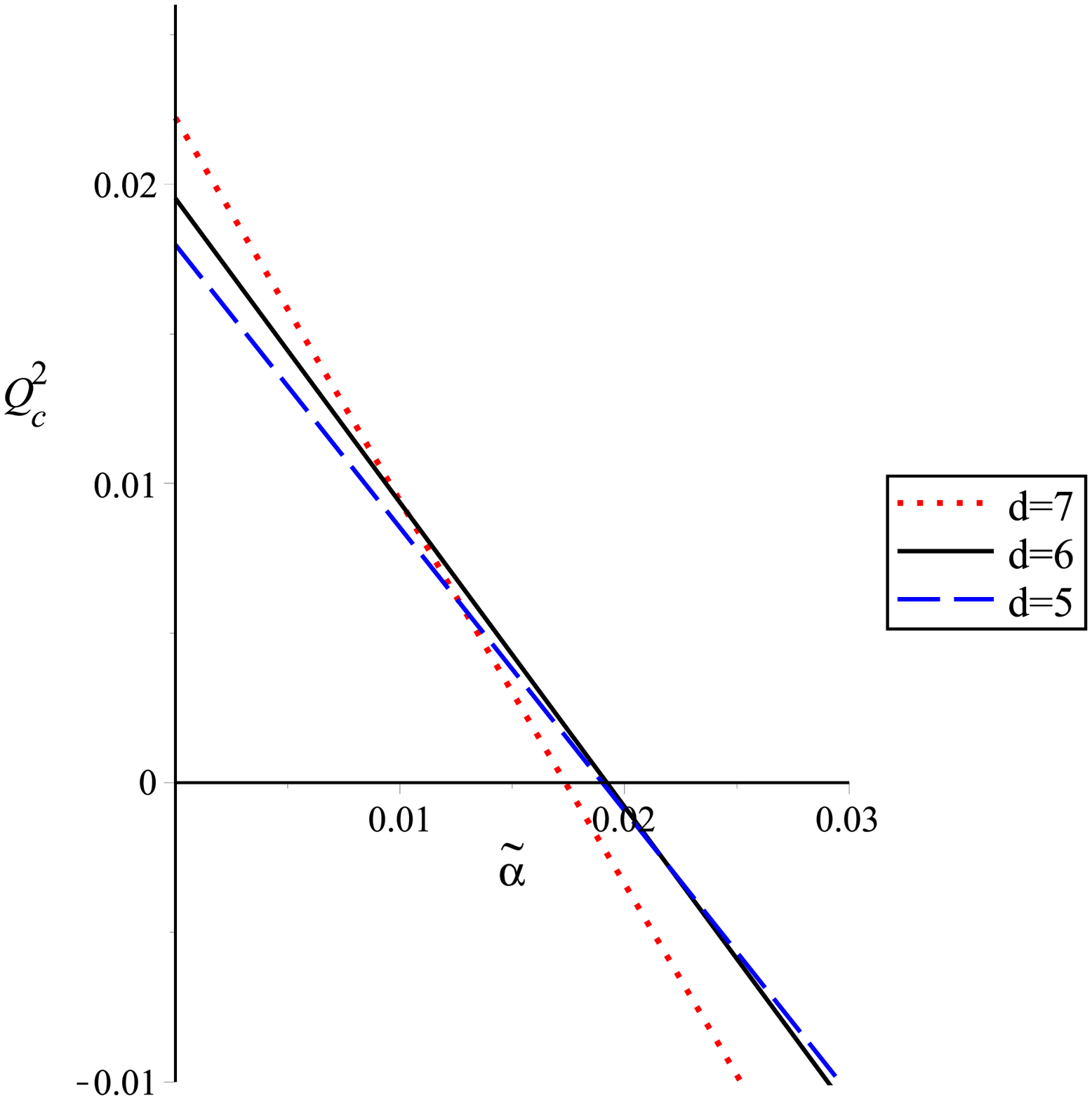}\label{Fig9b}}\caption{Critical
values in terms of $\tilde{\alpha}$ for GB black holes }
    \label{Fig9}
  \end{figure}

\begin{figure}
 \centering \subfigure[$\psi_c$ versus $\tilde{\alpha}$ ]
{\includegraphics[scale=0.3]{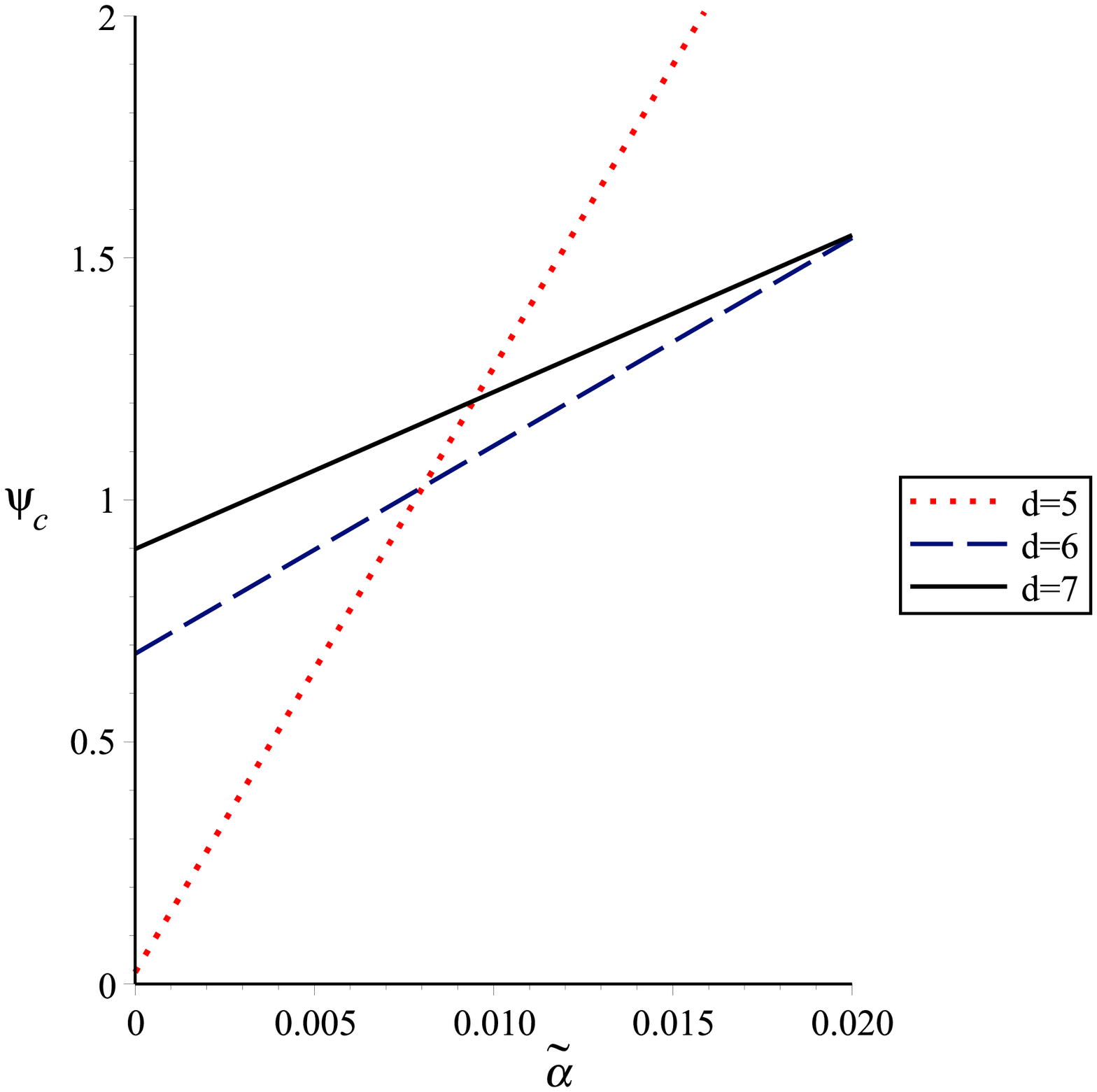}\label{Fig10a}}
  \hspace*{.1cm} \subfigure[$\rho_c$ versus $\tilde{\alpha}$
]{\includegraphics[scale=0.3]{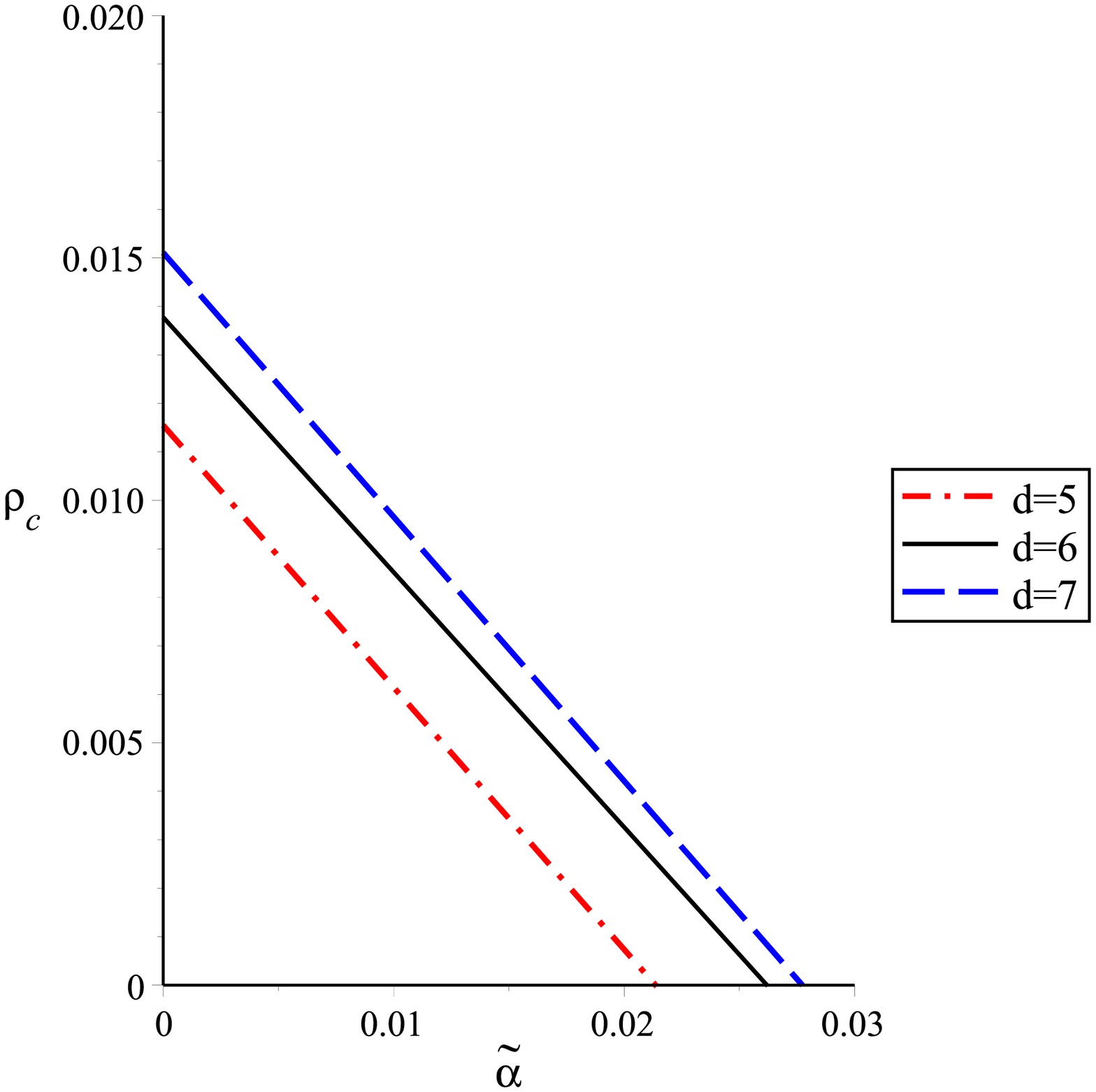}\label{Fig10b}}\caption{Critical
values in terms of $\tilde{\alpha}$ for GB black holes. }
    \label{Fig10}
  \end{figure}

\begin{figure}
 \centering \subfigure[ $d=5$ ]
{\includegraphics[scale=0.3]{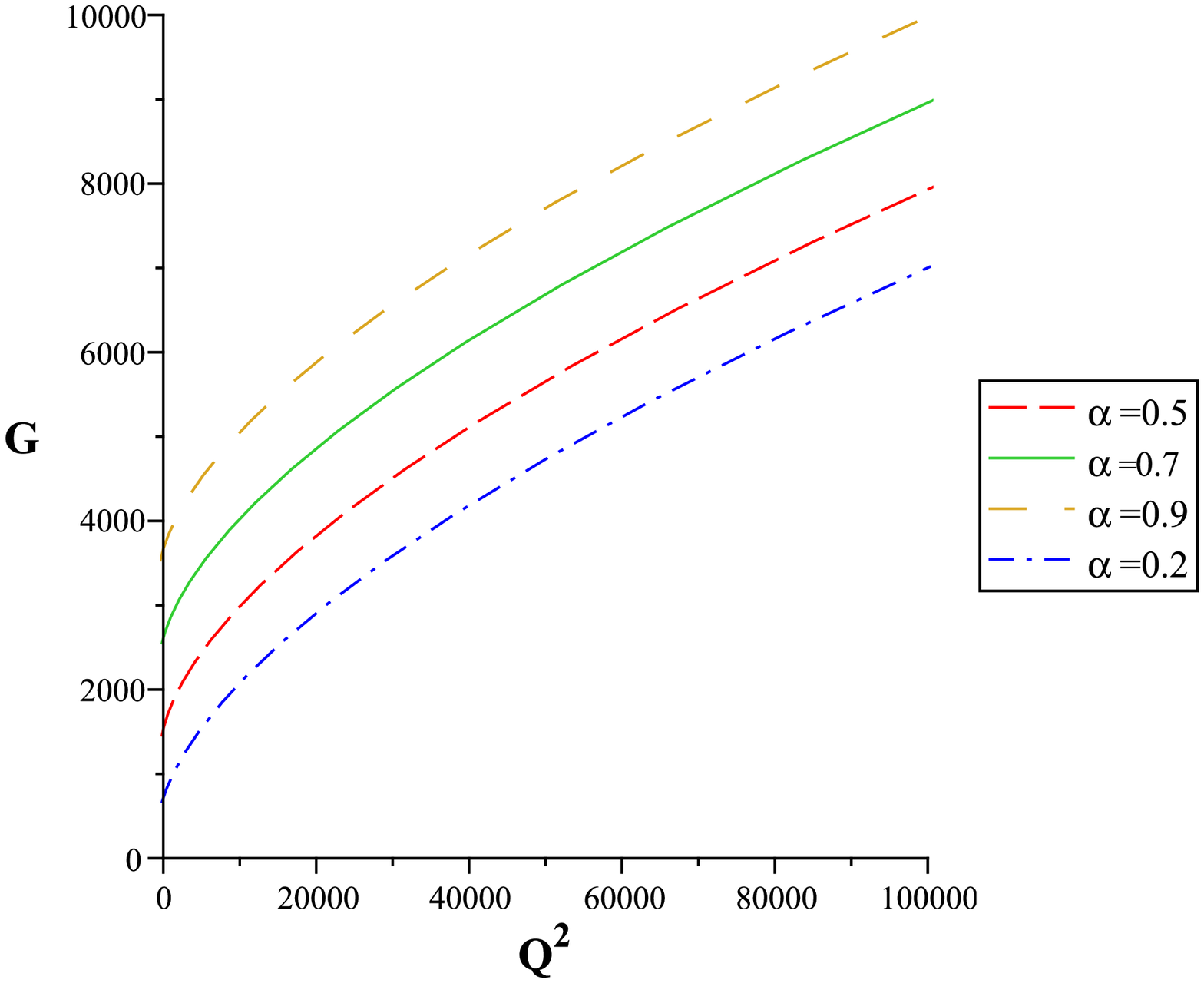}\label{Fig11a}}
  \hspace*{.1cm} \subfigure[ $d=6$]{\includegraphics[scale=0.3]{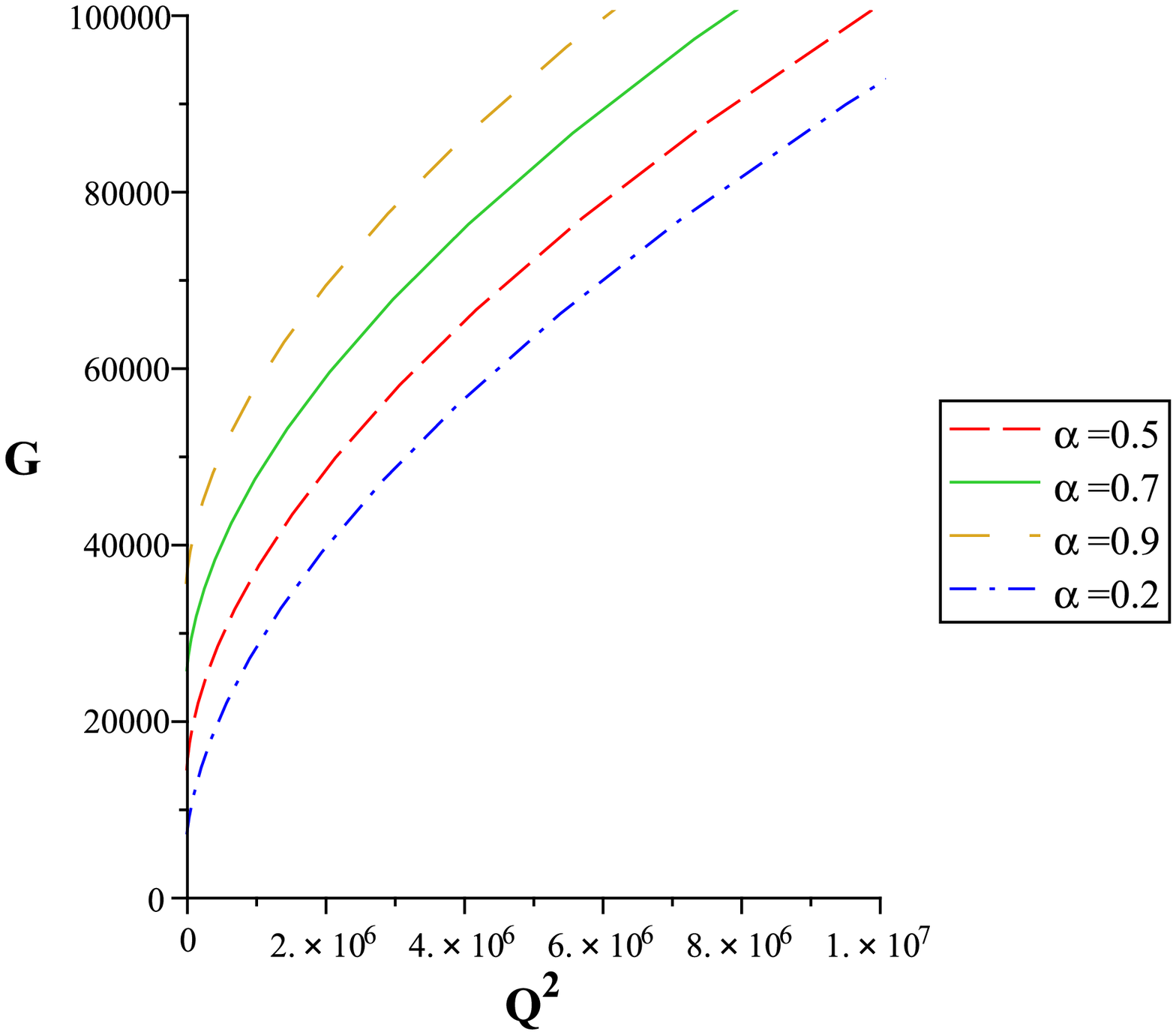}
  \label{Fig11.b}}\caption{$G-Q^2$ in Gauss-Bonnet black holes with $k=1$, $l=1$ and different values of $\tilde{\alpha}$.}
    \label{fig11}
  \end{figure}

It is worth studying the behavior of the critical values with
respect to $\tilde{\alpha}$,  which displayed in Figs. \ref{Fig9}
and \ref{Fig10}. We see from Fig. \ref{Fig9} that $T_c$ and
$Q^2_c$ decrease when $\tilde{\alpha}$ increases. In addition, the
large value of $\tilde{\alpha}$ leads to negative values for $T_c$
and $Q^2_c$. This implies that $\tilde{\alpha}$ should have an
upper bound which depends on the dimension of spacetime. On the
other hand, as one may see from Fig. \ref{Fig10} the values of
$\psi_c$ and $\rho_c$ increase with increasing $\tilde{\alpha}$
with no upper bound.
We also plot  the Gibbs free energy versus $Q^2$ for diffract
values of $\tilde{\alpha}$ for both $d=5$  and $d=6$ in Fig.
\ref{fig11} which show that all cases have upward trends.
\section{Summery and conclusion}
We have investigated the critical behavior of charged GB black
holes in AdS spaces via an alternative phase space. In this
approach, one can treat the square of the charge of the black hole
as a thermodynamic variable and fix the cosmological constant. It
is more reasonable to take the charge of the black hole as an
external variable which can vary, instead of the cosmological
constant which basically has a constant value. For example, one
may think that the charge of the black hole can change by
absorbing or emitting the charged particles. The advantages of
this new approach is that we do not need to extend the
thermodynamical phase space, in order to see the critical
behaviour of the system. It was argued that this new approach
admits a critical behavior for the black holes similar to the Van
der Waals liquid-gas system with the same critical exponents
provided one treat the square of the charge of the black hole
($Q^2$) as the thermodynamic variable \cite{Dehy}.

In this paper, we first generalized the method developed in
\cite{Dehy} to all higher dimensions by investigating the critical
behaviour of $d$-dimensional charged AdS black holes and treating
$Q^2$ as the thermodynamic variable and keeping $\Lambda$
constant. We found that the critical behaviour of the system in
$Q^2-\Psi$ plane and the critical exponents are similar to the the
Van der Waals fluid system. Then, we applied this new approach to
string inspired GB gravity. We found out that the phase transition
occurs only for the small values of GB coupling constant
($\tilde{\alpha}$). Besides, the critical quantities are
reasonable, provided $\tilde{\alpha}$ to be small and the topology
of the horizon is assumed spherical. We found that these black
holes may have one or two horizons for small $\tilde{\alpha}$.
Therefore, there is neither horizon nor phase transition for
larger value of the dilaton coupling constant $\tilde{\alpha}$. We
calculated the critical quantities such as $T_c$, $\Psi_c$, $Q_c$
and $\rho_c$ and the critical exponents and observed the critical
temperature and critical charge go to zero as $\tilde{\alpha}$
increases. Furthermore, we calculated the Gibbs free energy of the
system. The swallowtail shapes of the Gibbs diagrams show the
existence of first order phase transition in the system. Also the
zero order phase transition are not seen in the diagrams. Finally,
we calculated the critical exponents of the GB black holes in all
higher dimensions and observed that they are independent of the
details of the system and are the same as those of Van der Waals
fluid.
\acknowledgments{We thank Shiraz University Research Council. The
work of AS has been supported financially by Research Institute
for Astronomy and Astrophysics of Maragha, Iran.}

\end{document}